\documentclass[12pt]{article}
\usepackage{color}
\usepackage{amsfonts,amssymb,amsmath}
\usepackage{theorem,cite}
\theorembodyfont{\rmfamily}
\newtheorem{rmk}{Remark}[section]

\numberwithin{equation}{section}

\definecolor{brique}{rgb}{.9,.2,0}
\definecolor{blvert}{rgb}{0,.8,.85}
\definecolor{vertcl}{rgb}{0,1,.7}
\newcommand\vertcl[1]{\textcolor{vertcl}{#1}}
\newcommand\blvert[1]{\textcolor{blvert}{#1}}
\newcommand\brique[1]{\textcolor{brique}{#1}}
\def\lapth{
\begin{picture}(164,70)(0,-15)\thicklines
\put(0,0){\vertcl{\rule{20pt}{4pt}}}
\put(19,1){\vertcl{\line(1,3){23}}} 
\put(20,1){\vertcl{\line(1,3){23}}} 
\put(21,1){\vertcl{\line(1,3){23}}}
\put(22,1){\vertcl{\line(1,3){23}}}
\put(45,70){\vertcl{\line(1,-3){23}}} 
\put(44,70){\vertcl{\line(1,-3){23}}} 
\put(43,70){\vertcl{\line(1,-3){23}}}
\put(42,70){\vertcl{\line(1,-3){23}}}
\put(2,24){\vertcl{\rule{120pt}{4pt}}}
\put(65,0){\vertcl{\rule{60pt}{4pt}}}
\put(5,37){\Huge{\brique{\textbf{L}}}} 
\put(62,37){\Huge{\brique{\textbf{PTh}}}}
\put(12,-8){\blvert{\rule{92pt}{3.5pt}}}
\put(24,-15){\blvert{\rule{57pt}{3.5pt}}}
\put(36,-22){\blvert{\rule{30pt}{3.5pt}}}
\end{picture}
\raisebox{35pt}{
\begin{minipage}{320pt}\begin{center}
\textbf{Laboratoire d'Annecy-le-Vieux de Physique
Th\'eorique}\\[4ex]
website: \texttt{http://lappweb.in2p3.fr/lapth-2005/}
\end{center}
\end{minipage}}\\
\vspace{10pt}\quad \hrulefill\\
\vspace{10pt}}

\newcommand{\ben}{\begin{eqnarray}}
\newcommand{\een}{\end{eqnarray}}
\newcommand{\beno}{\begin{eqnarray*}}
\newcommand{\eeno}{\end{eqnarray*}}
\newcommand{\beq}{\begin{equation}}
\newcommand{\eeq}{\end{equation}}
\newcommand{\nonu}{\nonumber \\}

\newcommand{\hs}[1]{\hspace{#1 mm}}

\newcommand{\eps}{\epsilon}

\def\cA{{\cal A}}

          \def\cN{{\cal N}}          
\def\cP{{\cal P}}                    \def\cR{{\cal R}}
          \def\cT{{\cal T}}

        \def\cn{\mbox{\small{{\sc n}}}}     
\def\fD{{\mathfrak D}}

\def\fa{{\mathfrak a}}

\def\fm{{\mathfrak m}}
\def\fn{{\mathfrak n}}
\def\fr{{\mathfrak r}}
\def\ft{{\mathfrak t}}

\newcommand{\CC}{{\mathbb C}}

\newcommand{\II}{{\mathbb I}}

\newcommand{\RR}{\mbox{${\mathbb R}$}}

\newcommand{\ZZ}{{\mathbb Z}}

\newcommand{\prt}{\partial}

\newcommand{\wh}[1]{\widehat{#1}}
\newcommand{\wt}[1]{\widetilde{#1}}

\newcommand{\mb}[1]{\hs{4}\mbox{#1}\hs{4}}

\newcommand{\half}{\frac{1}{2}}

\newcommand{\arctanh}{\mbox{arctanh}}

\begin{document}
\renewcommand{\thefootnote}{\fnsymbol{footnote}}
\newpage
\pagestyle{empty}
\setcounter{page}{0}
%%%%%%%%%%%%%%%%%%%%%%%%%%%%%%%%
%%%%%  HEADINGS POUR DRAFT  %%%%%%

\markright{\today\dotfill DRAFT\dotfill }
% \pagestyle{myheadings}

%\section{titre}
\hspace{-1cm}\lapth

\vfill

\begin{center}

{\LARGE  {\sffamily Quantum field theory on quantum graphs\\[1.2ex]
and application to their conductance
}}
\vfill
  
{\large E. Ragoucy\footnote{ragoucy@lapp.in2p3.fr}}\\[1.2ex] 
{\large\textit{Laboratoire de Physique Th{\'e}orique LAPTH}}
\\[1.2ex]
UMR 5108 
   du CNRS, associ{\'e}e {\`a} l'Universit{\'e} de Savoie\\
   LAPP, BP 110, F-74941  Annecy-le-Vieux Cedex, France. 
\end{center}
\vfill

\begin{abstract}
We construct a bosonic quantum field on a general quantum graph. 
Consistency of the construction leads to the calculation of the total 
scattering matrix of the graph. This matrix is equivalent to the one 
already proposed using generalized star product approach. We give 
several examples and show how they generalize some of the scattering matrices 
computed in the mathematical or condensed matter physics literature.

Then, we apply the construction for the calculation of the 
conductance of graphs, within a small distance approximation. The 
consistency of the approximation is proved by direct comparison with 
the exact calculation for the `tadpole' graph.
\end{abstract}

\vfill

\rightline{January 2009}
\rightline{LAPTH-1304/09}
\rightline{\tt arXiv:0901.2431 [hep-th]}

\newpage
\pagestyle{plain}
\setcounter{page}{1}
\renewcommand{\thefootnote}{\arabic{footnote}}
\setcounter{footnote}{0}
\section{Introduction}
Quantum graphs have been recently the subject of intense studies, 
both at the mathematical level, see e.g. 
\cite{ES,Harm,Kuch,Schra2,Schra3,Schra,Schra-mag,stargraph,conduc,bosostar,
boundstate,Schra-unp} 
and references therein, and for condensed matter 
physics applications in wires, see e.g. \cite{fish,Cham1,Cham2,das} and 
references therein, or chaos \cite{chaos}. 
These graphs appear to be a very good 
approximation for the modeling of quasi unidimensional systems, 
such as quantum or atomic wires (for a review see e.g. \cite{rev1,rev2}). 

In this paper, we show how to construct quantum fields on a general 
graph, starting from the knowledge of the scattering matrix at each 
vertex of this graph. The construction relies on the RT-algebra 
formalism and gives a way to compute the total scattering matrix 
associated to the graph. This total scattering matrix is equivalent to 
the one constructed using the generalized star product framework 
\cite{Schra2,Schra,Schra-unp}. Then, we apply the formalism to the explicit 
calculation of conductance for Tomonaga-Luttinger models for
specific graphs, such as tree graph, the loop, 
the tadpole and the triangle. 

The paper consists of two different parts. The first one (that 
contains sections \ref{sec:resu}, \ref{sec:sim2V}, \ref{sec:gen2V} 
and \ref{sec:genVgen}) deals with 
the formal aspects of the construction. 
If one assumes the results of this part,
one can directly read the second part 
(containing sections \ref{sec:scalinv} and \ref{sec:conduc}) that is 
focused on explicit calculations and examples. 

More specifically, we summarize in section  \ref{sec:resu} known 
results \cite{stargraph,conduc,bosostar} on quantum field theory on star graphs. In section 
\ref{sec:sim2V}, we show how to construct a bosonic quantum field on a 
graph consisting of two star graph linked by a single line. The 
construction is essentially based on the determination of 
its total scattering matrix. In section \ref{sec:gen2V}, we 
generalize the approach to the case where several lines are tied 
between the two star graphs, and in section \ref{sec:genVgen} we treat 
the general case of several star graphs linked by several lines. 
In section \ref{sec:scalinv}, we apply the previous results to the 
case of scale invariant scattering matrices. This allows us to recover 
results obtained both in mathematical physics literature 
\cite{Schra,Schra2,Schra-unp} and in condensed matter physics \cite{Cham1,Cham2}.
Finally, using the technics developed in \cite{conduc,boundstate}, we apply 
in section \ref{sec:conduc} our 
results to the calculation of conductance on graphs. The calculation 
is done in a short distance approximation. 
In the case of a tadpole graph, we compute the conductance exactly 
and show that the approximation is consistent with the exact 
calculation. An appendix is devoted to the proofs of 
the properties used in the paper. 

\section{Integrable field theory on star graphs\label{sec:resu}}
We summarize here the results developed in \cite{stargraph}--\cite{bosostar} 
for the construction of an integrable field theory on a star graph. 
The algebraic 
framework needed to define bosonic fields on star graph are the 
RT-algebras \cite{RTalg1,RTalg2}. These algebras are a generalization of ZF-algebras, 
themselves being a generalization of oscillator algebras. Indeed, if 
oscillator algebras are used to define free fields on, say, an infinite 
line, ZF algebras are adapted to define interacting fields on this 
line, while RT-algebras take into account the introduction of a 
defect on this line. 

\subsection{RT algebras}
\begin{figure}[htb]
\begin{center}
\begin{picture}(300,90) \thicklines
\put(40,38){$
\begin{array}{c} \fa_{1} \\[1.2pt] \fa_{2} \\[1.2pt] \fa_{3} 
\\[1.2pt] \vdots \\[1.2pt] \vdots 
\end{array}$}
\put(120,40){\vector(-1,0){35}}
\put(120,40){\vector(-2,1){35}}
\put(120,40){\vector(-2,-1){35}}
\put(120,40){\vector(-1,1){35}}
\put(120,40){\vector(-1,-1){35}}
\put(118,40){\circle*{7}}
\put(115,10){$S(p)$}
\put(118,40){\vector(1,0){35}}
\put(118,40){\vector(2,1){35}}
\put(118,40){\vector(2,-1){35}}
\put(118,40){\vector(1,1){35}}
\put(118,40){\vector(1,-1){35}}
\put(166,38){$
\begin{array}{c}  \vdots \\[1.2pt] \vdots \\[1.2pt] \vdots 
\\[1.2pt] \fa_{\fn-1} \\[1.2pt] \fa_{\fn} 
\end{array}$}
\end{picture}
\end{center}
\caption{Star graphs\label{fig:star} (the arrows 
indicate the orientation on the edges)}
\end{figure}
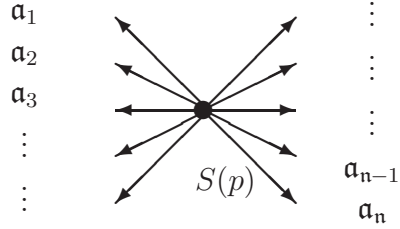

We present the RT algebra for a star graph with $\fn$ 
edges consisting in $\fn$ infinite half-lines (the edges) 
originating from the same point (the vertex), see figure \ref{fig:star}. 

To each edge $a=1,\ldots,\fn$, one associates oscillator-like 
generators $\{\fa_{a}(p),\,\fa_{a}^\dag(p)\}$ that deal with the field 
propagating on the edge. 
They are gathered in row and 
line vectors:
\begin{equation}
A(p) = \left(\begin{array}{c} \fa_{1}(p) \\ \fa_{2}(p) \\ \vdots \\ 
\fa_{\fn}(p) \end{array}\right)
\mb{and}
A^\dag(p) = \Big(\fa^\dag_{1}(p) \,,\, \fa^\dag_{2}(p) \,,\, \dots \,,\,
\fa^\dag_{\fn}(p) \Big)\,.
\end{equation}
It remains to give the (integrable) boundary condition at 
the vertex, i.e. the way the field connects between the different edges. 
This boundary condition is given by $\fn^2$ generators 
$s_{a_{1}a_{2}}(p)$, gathered in a matrix, the scattering matrix of 
the vertex: 
\begin{equation}
S(p) = \left(\begin{array}{cccc} 
s_{11}(p) & s_{12}(p) & \dots & s_{1\fn}(p) \\
s_{21}(p) & s_{22}(p) & \dots & s_{2\fn}(p) \\
\vdots &   & \ddots & \vdots \\
s_{\fn1}(p) & s_{\fn2}(p) & \dots & s_{\fn\fn}(p) 
\end{array}\right)\,.
\end{equation}
The RT-algebra is the unital algebra generated by\footnote{Strictly 
speaking, at the algebraic level, the RT algebra can be defined for 
$p\in\CC$. However, since $p$ is physically associated to an 
impulsion, we restrict ourself to real $p$'s.}
$\{\fa_{a}(p),\,\fa_{b}^\dag(p)\,,\, s_{ab}(p)\,,\ 
a\,,\,b=1,\ldots,\fn\,,\, p\in\RR\}$ submitted to
the relations:
\begin{eqnarray}
&&\fa_{a_{1}}(p_{1})\,\fa_{a_{2}}(p_{2}) 
- \fa_{a_{2}}(p_{2})\,\fa_{a_{1}}(p_{1}) =0
\,,\label{eq:RT1}\\
&&\fa^\dag_{a_{1}}(p_{1})\,\fa^\dag_{a_{2}}(p_{2}) 
- \fa^\dag_{a_{2}}(p_{2})\,\fa^\dag_{a_{1}}(p_{1}) =0
\,,\label{eq:RT2}\\
&&\fa_{a_{1}}(p_{1})\,\fa^\dag_{a_{2}}(p_{2}) 
- \fa^\dag_{a_{2}}(p_{2})\,\fa_{a_{1}}(p_{1}) =
2\pi\Big(\delta(p_{1}-p_{2})\,\delta_{a_{1}a_{2}} + 
\delta(p_{1}+p_{2})\,s_{a_{1}a_{2}}(p_{1})\Big)
\qquad \label{eq:RT3}
\end{eqnarray}
and the boundary condition
\begin{equation}
A(p) = S(p)\, A(-p)\mb{and} A^\dag(p) =  A^\dag(-p) \,S(-p)\,.
\label{eq:bound}
\end{equation}

The RT algebra admits an anti-automorphism (written for $p\in\RR$)
\beq
\fa_{a}(p)\ \to\ \fa^\dag_{a}(p) \mb{;} \fa^\dag_{a}(p)\ \to\ \fa_{a}(p)
\mb{and} s_{a_{1}a_{2}}(p)\ \to\ s_{a_{2}a_{1}}(-p)
\eeq
which is identified with the hermitian conjugation.

There are two consistency relations coming from relation 
(\ref{eq:bound}). For $p\in\RR$, they read:
\begin{eqnarray}
S(p)\, S(-p) =\II \label{cons}\\
S^\dag(p)=S(-p) \label{herm}
\end{eqnarray}
One recognizes in (\ref{herm}) Hermitian analycity
for the scattering matrix $S(p)$. Together with the consistency 
relation (\ref{cons}), it implies unitarity of the scattering 
matrix:
\beq
S(p)\,S^\dag(p) =\II\,. \label{unit}
\eeq
Below, we will decompose the scattering matrix into block 
submatrices:
\begin{equation}
\displaystyle
 S(p)=\left(\begin{array}{c|c}
S_{11}(p) &  S_{12}(p)\\ \hline
S_{21}(p) & S_{22}(p)
\end{array}\right)\,.
\end{equation}
Within this decomposition, the consistency relation (\ref{cons}) 
recasts into four equations:
\begin{eqnarray}
S_{11}(p)\, S_{11}(-p) +S_{12}(p)\, S_{21}(-p) =\II 
&;& S_{11}(p)\, S_{12}(-p) +S_{12}(p)\, S_{22}(-p) =0 
\qquad
\label{cons1}\\
S_{21}(p)\, S_{11}(-p) +S_{22}(p)\, S_{21}(-p) =0
&;&S_{22}(p)\, S_{22}(-p) +S_{21}(p)\, S_{12}(-p) =\II 
\label{cons2}
\end{eqnarray}

\subsection{Quantum field on star graph}
A massless bosonic field on the star graph is constructed from the 
RT-algebra generators as:
\begin{equation}
\phi_{a}(x,t)=\int_{-\infty}^\infty \frac{dp}{2\pi}
\frac{1}{\sqrt{2|p|}} \Big\{ e^{-i(|p|t-px)}\,\fa_{a}(p) + 
e^{i(|p|t-px)}\,\fa^{\dag}_{a}(p) \Big\}
\mb{}a=1,2,\ldots,\fn\,.
\label{eq:freefield}
\end{equation}
In expression (\ref{eq:freefield}), $x\geq0$ is the 
distance on the edge $a$ on which the 
field propagates, with 
origin at the vertex. Using the relations (\ref{eq:RT1})-(\ref{eq:RT3})
and (\ref{eq:bound}), 
it can be shown that the field $\phi$ has 
canonical equal time commutation on each edge:
\ben
&&\big[ \phi_{a_{1}}(x_{1},0)\,,\,\phi_{a_{2}}(x_{2},0)\big] =0
\mb{and}\big[ (\prt_{t}\phi_{a_{1}})(x_{1},0)\,,\,\phi_{a_{2}}(x_{2},0)\big] =
-i\,\delta_{a_{1}a_{2}}\,\delta(x_{1}-x_{2})\,,
\nonu
&& x_{1}\,,\,x_{2}>0\,,\,a_{1},\,a_{2}=1,\ldots,\fn\,;
\label{eq:cano}
\een
obeys the equation of motion:
\beq
\big(\prt_{t}^2-\prt_{x}^2\big)\phi_{a}(x,t) = 0
\mb{,} x>0\,,\  a=1,2,\ldots,\fn\,;
\label{eq:mvt}
\eeq
and some boundary condition which depends on the form of $S(p)$.
 When the scattering matrix takes the 
form $S(p)= - (B+ip\,C)^{-1}\,(B-ip\,C)$, where $B$ and $C$ are real 
matrices such that $B\,C^t=C\,B^t$, this boundary condition reads:
\beq
\sum_{b=1}^{\fn}\Big(B_{ab}\,\phi_{b}(0,t) +
C_{ba}(\prt_{x}\phi_{b})(0,t) \Big)= 0
\mb{,} t\in\RR\,,\  a=1,2,\ldots,\fn\,.
\eeq
Equation (\ref{eq:freefield}) corresponds to a planar wave 
decomposition of the field $\phi_{a}(x,t)$. We will
 call $\fa_{a}(p)$ the mode (or the oscillator) on the 
edge $a$ (with momentum $p$).

\paragraph{Choice of the origin on each edge} 
In the following, we will need to change the origin of coordinate on 
some edges. This amounts to change the form of the scattering matrix. 
Indeed, from the form (\ref{eq:freefield}), it is clear that
 a shift $x\,\to\,x+d$ is equivalent to the transformation
\beq
\fa_{a}(p)\ \to\ e^{ipd}\,\fa_{a}(p) \mb{and}
\fa^{\dag}_{a}(p)\ \to\ e^{-ipd}\,\fa^{\dag}_{a}(p)\,. 
\eeq
This transformation does not modify the relations 
(\ref{eq:RT1}) and (\ref{eq:RT2}), but it does affect the  
scattering matrix in (\ref{eq:RT3}). For general shifts of $x\,\to\,x+d_{a}$,
 on the edge $a=1,\ldots,\fn$, the scattering matrix will be changed 
as
\begin{equation}
S(p)\ \to\ W(p)\,S(p)\,W(p) \mb{with} 
W(p)=\mbox{diag}(e^{ipd_{1}},\ldots,e^{ipd_{\fn}})\,.
\label{eq:dist}
\end{equation}
Remark that since $S(p)$ obeys the consistency and unitarity relations 
(\ref{cons})-(\ref{unit}), the transformation (\ref{eq:dist}) does 
not change the properties (\ref{cons})-(\ref{unit}) of the scattering 
matrix.

In the same way, a change of orientation on the edges will correspond 
to a transformation: $\fa(p)\ \to\ \fa(-p)$ in the boundary 
condition.

\subsection{Star graphs as building blocks for quantum wires}
In the following we shall construct integrable quantum field theory 
on a general quantum wire. It should be clear that the star graphs can 
be considered as building blocks for such general wire, in the same 
way the single defect on a line underlies the construction for several 
defects on the line \cite{multidef}. In both cases, the 
above construction applies \textit{locally}, around 
each vertex. The scattering matrices attached to each vertices will be 
called local. They are part of the quantum graph data.
 What remains to do is to connect these star graphs, i.e. 
identify the field on connecting edges between two star graphs. We 
will see that this physical identification is sufficient to determine the 
`internal' modes (i.e. the generators $\{\fa_{a}(p),\fa^\dag_{a}(p)\}$ of a 
connecting edge $a$ between two vertices) in terms of the `external' 
modes. It allows also to construct the global scattering matrix, that 
relates the `external' modes $\{\fa_{b}(p),\fa^\dag_{b}(p)\}$ (on 
external edges) through a relation of the type (\ref{eq:bound}). This 
natural 
identification (which leads to a purely algebraic calculation) appears 
to be 
equivalent to the one introduced in \cite{Schra} in analyzing the 
Schr\"odinger operator on graphs. In both cases, one needs to `glue' 
star graphs together, either using a generalized star-product 
\cite{Schra}, or through identification of the bosonic modes 
propagating on the connecting edge(s).

\section{Simple gluing of two vertices\label{sec:sim2V}}
\subsection{General presentation}
\begin{figure}[htb]
\begin{center}
\begin{picture}(300,90) \thicklines
\put(0,38){$n-1\ \left\{
\begin{array}{c} \fa_{1} \\[1.2pt] \fa_{2} \\[1.2pt] \fa_{3} \\[1.2pt] \vdots 
\\[1.2pt] \fa_{n-1} 
\end{array}\right. $}
\put(120,40){\vector(-1,0){35}}
\put(120,40){\vector(-2,1){35}}
\put(120,40){\vector(-2,-1){35}}
\put(120,40){\vector(-1,1){35}}
\put(120,40){\vector(-1,-1){35}}
\put(118,40){\circle*{7}}
\put(118,25){$S(p)$}
\put(120,40){\vector(1,0){20}}
\multiput(140,40)(10,0){4}{\line(1,0){4}}
\put(155,50){$\fa_{\fn}$}
\put(200,40){\line(-1,0){20}}
\put(190,40){\vector(1,0){5}}
\put(180,25){$\Sigma(p)$}
\put(202,40){\circle*{7}}
\put(200,40){\vector(1,0){35}}
\put(200,40){\vector(2,1){35}}
\put(200,40){\vector(2,-1){35}}
\put(200,40){\vector(1,1){35}}
\put(200,40){\vector(1,-1){35}}
\put(240,38){$\left.
\begin{array}{c} \fa_{\fn+1} \\[1.2pt] \fa_{\fn+2} \\[1.2pt] \fa_{\fn+3} \\[1.2pt] \vdots 
\\[1.2pt] \fa_{\fn+\fm-1} 
\end{array}\right\}\ \fm-1$}
\end{picture}
\end{center}
\caption{Simple gluing of two vertices
 (The arrows indicate the orientation of the edge)
 \label{fig:sim2V}}
\end{figure}
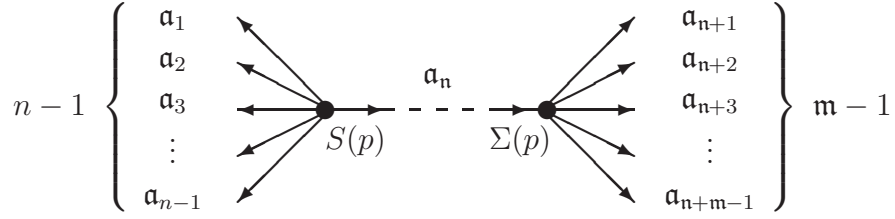

We consider two star graphs with $\fn$ and $\fm$ edges respectively,
that are linked 
by one edge. We want to construct the quantum field on this graph. The 
basic idea is that \textit{locally} around each vertex, the bosonic field 
should be the same as the one for the corresponding star graph. Then, 
one should connect the two construction via the connecting edge, 
where the two fields should correspond.
 We call this 
procedure the `gluing' of the two vertices. It is drawn in figure 
\ref{fig:sim2V}.

The local $S$ matrices are denoted
\begin{equation}
S(p)=\left(\begin{array}{ccc}
s_{11}(p) & \ldots & s_{1\fn}(p) \\
\vdots & & \vdots \\
s_{\fn1}(p) & \ldots & s_{\fn\fn}(p)
\end{array}\right)
\mb{and}
\Sigma(p)=\left(\begin{array}{ccc}
\sigma_{11}(p) & \ldots & \sigma_{1\fm}(p) \\
\vdots & & \vdots \\
\sigma_{\fm1}(p) & \ldots & \sigma_{\fm\fm}(p)
\end{array}\right)\,.
\end{equation}

The line that links the two vertices is 
denoted $\fn$ in $S(p)$ and 1 in 
$\Sigma(p)$. For each edge 
$a\neq \fn$ the origin is chosen to be at the vertex to which the edge 
belongs. For the edge $\fn$, the origin is chosen at the vertex 
described by $S(p)$, so that $S(p)$ is the `true' local scattering 
matrix of the vertex, while $\Sigma(p)$ is related to the `true' 
local scattering matrix $\Sigma_{0}(p)$ by 
\begin{equation}
\Sigma(p) = W(p)\,\Sigma_{0}(p)\,W(p) \mb{with} 
W(p)=\mbox{diag}(e^{ipd_{\fn}},1,1,\ldots,1)
\end{equation}
where $d_{\fn}$ is the distance between the two vertices (measured on 
the edge $\fn$).

The boundary conditions on each vertex are local, and hence take the 
form 
\begin{equation}
\wh A(p) = S(p)\,\wh A(-p) \mb{and} \wh B(p)=\Sigma(p)\,\wh B(-p)
\label{eq:bound-simple}
\end{equation}
where
\begin{equation}
\wh A(p)=\left(\begin{array}{c} 
\fa_{1}(p) \\ \fa_{2}(p) \\ \vdots \\ \fa_{\fn}(p)
\end{array}\right)
\mb{and}
\wh B(p)=\left(\begin{array}{c} 
\fa_{\fn}(-p) \\ \fa_{\fn+1}(p) \\ \vdots \\ \fa_{\fn+\fm-1}(p)
\end{array}\right)\,.
\end{equation}
Since the mode $\fa_{\fn}(p)$ is common to $A(p)$ and $B(p)$, one can 
eliminate it from the system. In other word, the field on the 
`inner line' of the graph is constructed from the modes on the 
outer lines. In order to do this calculation, 
we single out $\fa_{n}(p)$:
\begin{equation}
\wh A(p)=\left(\begin{array}{c} 
A(p) \\[1.2ex] \fa_{\fn}(p)
\end{array}\right)
\mb{and}
\wh B(p)=\left(\begin{array}{c} 
\fa_{\fn}(-p) \\[1.2ex] B(p)
\end{array}\right)\,,
\end{equation}
where we have introduced
\begin{equation}
\displaystyle A(p)=\left(\begin{array}{c} 
\fa_{1}(p) \\ \vdots \\ \fa_{\fn-1}(p)
\end{array}\right)
\mb{and}
 B(p)=\left(\begin{array}{c} 
 \fa_{\fn+1}(p) \\ \vdots \\ \fa_{\fn+\fm-1}(p)
\end{array}\right) \,.
\end{equation}
We apply the same decomposition to the matrices $S(p)$ and $\Sigma(p)$:
\begin{equation}
\begin{array}{ll}
\displaystyle
 S_{11}(p)=\left(\begin{array}{ccc}
s_{11}(p) & \ldots & s_{1,\fn-1}(p) \\
\vdots & & \vdots \\
s_{\fn-1,1}(p) & \ldots & s_{\fn-1,\fn-1}(p)
\end{array}\right)
&\mb{;}
\Sigma_{22}(p)=\left(\begin{array}{ccc}
\sigma_{22}(p) & \ldots & \sigma_{2\fm}(p) \\
\vdots & & \vdots \\
\sigma_{\fm2}(p) & \ldots & \sigma_{\fm\fm}(p)
\end{array}\right)
\\[5.4ex]
\displaystyle
S_{21}(p) = \Big(s_{\fn1}(p),\ldots,s_{\fn,\fn-1}(p)\Big)
&\mb{;} \Sigma_{12}(p) = \Big(\sigma_{12}(p),\ldots,\sigma_{1\fm}(p)\Big)
\\[2.1ex]
\displaystyle
S_{12}(p) = \left(\begin{array}{c}
s_{1\fn}(p) \\ \vdots \\ s_{\fn-1,\fn}(p) \end{array}\right)
&\mb{;} \Sigma_{21}(p) = \left(\begin{array}{c}
\sigma_{21}(p) \\ \vdots \\ \sigma_{\fm1}(p) \end{array}\right)
\end{array}
\end{equation}
so that
\begin{equation}
S(p)=\left(\begin{array}{cc}
S_{11}(p)  & S_{12}(p) \\[1.2ex]
S_{21}(p) & s_{\fn\fn}(p)
\end{array}\right)
\qquad\qquad
\Sigma(p)=\left(\begin{array}{cc}
\sigma_{11}(p)  &   \Sigma_{12}(p) \\[1.2ex]
\Sigma_{21}(p) & \Sigma_{22}(p)
\end{array}\right)\,.
\end{equation}

With these splittings, the boundary conditions (\ref{eq:bound-simple}) 
recast as
\begin{eqnarray}
A(p) &=& S_{11}(p)\,A(-p)+S_{12}(p)\,\fa_{\fn}(-p) 
\\
\fa_{\fn}(p) &=& S_{21}(p)\,A(-p)+s_{\fn\fn}(p)\,\fa_{\fn}(-p)
\label{eq:an-a}\\
B(p) &=& \Sigma_{22}(p)\,B(-p)+\Sigma_{21}(p)\,\fa_{\fn}(p) 
\\
\fa_{\fn}(-p) &=& \Sigma_{12}(p)\,B(-p)+\sigma_{11}(p)\,\fa_{\fn}(p)
\label{eq:an-b}
\end{eqnarray}
Equations (\ref{eq:an-a}) and (\ref{eq:an-b}) allow us to express $\fa_{n}(p)$ 
in term of $A(p)$ and $B(p)$
\begin{equation}
\fa_{\fn}(p) = \frac{1}{1-\sigma_{11}(p)\,s_{\fn\fn}(p)}\,\Big(
S_{21}(p)\,A(-p) + s_{\fn\fn}(p)\,\Sigma_{12}(p)\,B(-p) \Big)
\label{eq:inner}
\end{equation}
together with a consistency relation
\ben
&& S_{21}(p)\,A(-p) + s_{\fn\fn}(p)\,\Sigma_{12}(p)\,B(-p) 
= \nonu
&&\qquad\qquad =
\frac{1-\sigma_{11}(p)\,s_{\fn\fn}(p)}{1-\sigma_{11}(-p)\,s_{\fn\fn}(-p)}\,
\Big(
\sigma_{11}(-p)\,S_{21}(-p)\,A(p) + \Sigma_{12}(-p)\,B(p) \Big)\,.
\qquad
\een
This consistency relation is automatically satisfied if 
$S(p)$ and $\Sigma(p)$ obey the consistency relation (\ref{cons}), see 
proof in appendix \ref{app:cons} for a more general case.
Then, defining 
\begin{equation}
\cA(p)=\left(\begin{array}{c} A(p) \\ B(p) \end{array}\right)
\end{equation}
we recast the two remaining relations as 
\begin{equation}
\cA(p)=S_{tot}(p)\,\cA(-p)
\end{equation}
 with 
\begin{equation}
S_{tot}(p) = \left(\begin{array}{cc}
\displaystyle
S_{11}(p)
+\frac{\sigma_{11}(p)\,S_{12}(p)\,S_{21}(p)}{1-\sigma_{11}(p)\,s_{\fn\fn}(p)}\quad
& \displaystyle
\frac{S_{12}(p)\,\Sigma_{12}(p)}{1-\sigma_{11}(p)\,s_{\fn\fn}(p)}
\\[3.1ex]
\displaystyle
\frac{\Sigma_{21}(p)\,S_{21}(p)}{1-\sigma_{11}(p)\,s_{\fn\fn}(p)}
&\displaystyle
\Sigma_{22}(p)
+\frac{s_{\fn\fn}(p)\,\Sigma_{21}(p)\,\Sigma_{12}(p)}{1-\sigma_{11}(p)\,s_{\fn\fn}(p)}
\end{array}\right)\,.
\label{eq:sim2V}
\end{equation}
One can show that  if $S(p)$ and $\Sigma(p)$ obey the consistency 
relation (\ref{cons}), then so does $S_{tot}(p)$. In the same way, 
the unitarity relation (\ref{unit}) for the matrices 
$S(p)$ and $\Sigma(p)$ implies unitarity for the matrix $S_{tot}(p)$.

The bosonic quantum field $\phi_{a}(x,t)$ keeps the form 
(\ref{eq:freefield}). Since the total scattering matrix obeys 
relations (\ref{herm}) and (\ref{cons}), the field $\phi_{a}(x,t)$ on external edges ($a\neq\fn$), it still 
obeys relations (\ref{eq:cano}) and (\ref{eq:mvt}). However, on the edge $\fn$, one 
has to replace the generators $\{\fa_{\fn}(p),\,\fa_{\fn}^\dag(p)\}$ 
by their expression (\ref{eq:inner}), and it is not ensured that 
$\phi_{\fn}(x,t)$ is canonical.

Remark that the scattering matrix can be rewritten as 
\ben
S_{tot}(p) &=& \left(\begin{array}{cc}
\displaystyle
S_{11}(p) & 0
\\[3.1ex]
0 &\displaystyle \Sigma_{22}(p)
\end{array}\right) \nonumber\\[1.2ex]
&&
+\frac{1}{1-\sigma_{11}(p)\,s_{\fn\fn}(p)}\,
\left(\begin{array}{cc}
\displaystyle
{\sigma_{11}(p)\,S_{12}(p)\,S_{21}(p)}
& \displaystyle
{S_{12}(p)\,\Sigma_{12}(p)}
\\[3.1ex]
\displaystyle
{\Sigma_{21}(p)\,S_{21}(p)}
&\displaystyle
{s_{\fn\fn}(p)\,\Sigma_{21}(p)\,\Sigma_{12}(p)}
\end{array}\right)\,.\quad
\label{eq:sim2Vbis}
\een
The first term in (\ref{eq:sim2Vbis}) corresponds to the scattering 
matrix of the two local vertices without interaction (i.e. when the 
edge $\fn$ is removed), while the second 
term is the `perturbation' due to the link through $\fa_{n}(p)$. Let us 
stress that only $S_{tot}(p)$ is unitary.

\begin{rmk} \label{rmk:gluing}
In the limit of a vanishing distance between the vertices, 
$d_{\fn}\,\to\,0$,
the gluing of scattering matrices can be viewed as 
a recursive process to build higher dimensionnal scattering matrices, starting 
from low dimensionnal ones. The process ensures unitarity of the 
final matrix when original ones are.
\end{rmk}

\subsection{Example 1: the `tree graph'\label{sec:tree}}
\begin{figure}[htb]
\begin{center}
\begin{picture}(300,90) \thicklines
\put(70,70){$\fa_{1}$}
\put(70,5){$\fa_{2}$}
\put(120,40){\vector(-1,1){35}}
\put(120,40){\vector(-1,-1){35}}
\put(118,40){\circle*{7}}
\put(118,25){$S(p)$}
\put(120,40){\vector(1,0){20}}
\multiput(140,40)(10,0){4}{\line(1,0){4}}
\put(155,50){$\fa_{3}$}
\put(200,40){\line(-1,0){20}}
\put(190,40){\vector(1,0){5}}
\put(180,25){$\Sigma(p)$}
\put(202,40){\circle*{7}}
\put(200,40){\vector(1,1){35}}
\put(200,40){\vector(1,-1){35}}
\put(240,70){$\fa_{4}$}
\put(240,5){$\fa_{5}$}
\end{picture}
\end{center}
\caption{The tree graph\label{fig:tree}}
\end{figure}
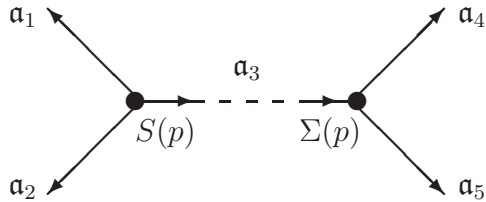

As an example we consider the gluing along one edge 
of two vertices with three edges. In this way, we construct the 
scattering matrix for a vertex with four edges, that we call a `tree 
graph' for obvious particle physics reasons, see figure \ref{fig:tree}. This gluing is the 
simplest example of the recursive process mentioned in remark 
\ref{rmk:gluing}. 

The decomposition of the local $S$ matrices reads
\begin{equation}
S(p) = \left(\begin{array}{cc|c} 
s_{11}(p) & s_{12}(p) & s_{13}(p) \\
s_{21}(p) & s_{22}(p) & s_{23}(p) \\
\hline
s_{31}(p) & s_{32}(p) & s_{33}(p) 
\end{array}\right)
\mb{and}
\Sigma(p) = \left(\begin{array}{c|cc} 
\sigma_{11}(p) & \sigma_{12}(p) & \sigma_{13}(p) \\
\hline
\sigma_{21}(p) & \sigma_{22}(p) & \sigma_{23}(p) \\
\sigma_{31}(p) & \sigma_{32}(p) & \sigma_{33}(p) 
\end{array}\right)
\end{equation}
where the lines indicate the block decomposition of the matrices.
The local boundary conditions have the form
\begin{equation}
\left(\begin{array}{c} \fa_{1}(p) \\ \fa_{2}(p) \\ \fa_{3}(p) 
\end{array}\right)\ =\ S(p)\, 
\left(\begin{array}{c} \fa_{1}(-p) \\ \fa_{2}(-p) \\ \fa_{3}(-p) 
\end{array}\right)
\mb{and}
\left(\begin{array}{c} \fa_{3}(-p) \\ \fa_{4}(p) \\ \fa_{5}(p) 
\end{array}\right)\ =\ \Sigma(p)\, 
\left(\begin{array}{c} \fa_{3}(p) \\ \fa_{4}(-p) \\ \fa_{5}(-p) 
\end{array}\right)\,.
\end{equation}

We will focus on identical local vertices. This does \underline{not}
mean that the local matrices $S(p)$ and $\Sigma(p)$ are identical, 
because of the different labeling 
and orientation of the edges on the total graph, and also because of the choice of the origin 
on the connecting edge. 
The local scattering matrices rather obey
\begin{equation}
\Sigma(p)=W(p)\,P\,S(-p)\,P^{-1}\,W(p)
\mb{with} P=\left(\begin{array}{ccc} 
0 & 0 & 1 \\ 1 & 0 & 0 \\ 0 & 1 & 0
\end{array}\right)
\ ;\ W(p)=\mbox{diag}(e^{ipd},1,1)
\label{identical}
\end{equation}
where $d$ is the distance between the two vertices. In 
(\ref{identical}), $P$ rotates the $S$-matrix according to
 the labelling of the edges, while $W(p)$ implements the shift 
of the origin, according to (\ref{eq:dist}).
It leads to
\begin{equation}
\Sigma(p) = \left(\begin{array}{c|cc} 
s_{33}(-p)\,e^{2ipd} & s_{31}(-p)\,e^{ipd} & s_{32}(-p)\,e^{ipd} 
\\[1.ex]
\hline\rule{0ex}{2.4ex}
s_{13}(-p)\,e^{ipd} & s_{11}(-p) & s_{12}(-p) \\
s_{23}(-p)\,e^{ipd} & s_{21}(-p) & s_{22}(-p) 
\end{array}\right)\,.
\end{equation}
Using this expression and the consistency relation, one rewrites 
(\ref{eq:sim2V}) as
\begin{eqnarray}
S_{tot}(p) &=& \left(\begin{array}{cc} S_{11}(p) & 0 \\
0 & S_{11}(-p) \end{array}\right)+
\frac{e^{2ipd}}{\cN(p)}\,
\left(\begin{array}{cc} s_{33}(-p)\,M(p,p) & 
e^{-ipd}\,M(p,-p) \\[1.2ex]
e^{-ipd}\,M(-p,p) &
s_{33}(p)\,M(-p,-p) 
\end{array}\right)\nonumber\\[1.2ex]
\cN(p) &=& 1-e^{2ipd}\,s_{33}(p)\,s_{33}(-p)
\label{eq:tree-gen}
\end{eqnarray}
where we have introduced the submatrix:
\begin{eqnarray}
M(p,q) &=& S_{12}(p)\cdot S_{21}(q)\,=\,
\left(\begin{array}{cc} 
s_{13}(p)\,s_{31}(q)\quad & 
s_{13}(p)\,s_{32}(q) \\[2.1ex]
s_{23}(p)\,s_{31}(q) & 
s_{23}(p)\,s_{32}(q) 
\end{array}\right)\,.
\end{eqnarray}
The boundary condition for the total tree graph reads
\begin{equation}
\left(\begin{array}{c} \fa_{1}(p) \\ \fa_{2}(p) \\ \fa_{4}(p) \\ \fa_{5}(p) 
\end{array}\right)\ =\ S_{tot}(p)\, 
\left(\begin{array}{c} \fa_{1}(-p) \\ \fa_{2}(-p) \\ \fa_{4}(-p) \\ \fa_{5}(-p) 
\end{array}\right)
\end{equation}
and the `inner mode' $\fa_{3}(p)$ is expressed in terms of the `outer 
modes' as
\ben
\fa_{3}(p) &=& \frac{1}{\cN(p)}\,\Big\{
s_{31}(p)\,\fa_{1}(-p)+s_{32}(p)\,\fa_{2}(-p)\nonu
&& \qquad\qquad+
e^{ipd}\,s_{33}(p)\Big(s_{31}(-p)\,\fa_{4}(-p)+s_{32}(-p)\,\fa_{5}(-p)
\Big)\Big\}\,.
\een

\section{General gluing of two vertices\label{sec:gen2V}}
\begin{figure}[htb]
\begin{center}
\begin{picture}(400,100) \thicklines
\put(30,48){$\fn\ \left\{
\begin{array}{c} \fa_{1} \\[1.2pt] \fa_{2} \\[1.2pt] \fa_{3} 
\\[1.2pt] \vdots \\[1.2pt] \fa_{\fn} 
\end{array}\right. $}
\put(120,50){\vector(-1,0){35}}
\put(120,50){\vector(-2,1){35}}
\put(120,50){\vector(-2,-1){35}}
\put(120,50){\vector(-1,1){35}}
\put(120,50){\vector(-1,-1){35}}
\put(110,70){$S(p)$}
\put(120,50){\circle*{7}}
\put(120,50){\vector(1,0){35}}
\put(120,50){\vector(2,1){35}}
\put(120,50){\vector(2,-1){35}}
\put(120,50){\vector(1,1){35}}
\put(120,50){\vector(1,-1){35}}
\put(180,89){$\fa_{\fn+1}$}
\multiput(158,85)(8,0){8}{\line(1,0){4}}
\put(178,73){$\fa_{\fn+2}$}
\multiput(158,68)(8,0){8}{\line(1,0){4}}
\put(178,55){$\fa_{\fn+3}$}
\multiput(158,50)(8,0){8}{\line(1,0){4}}
\put(180,35){$\quad \vdots$}
\multiput(158,32)(8,0){8}{\line(1,0){4}}
\put(175,18){$\fa_{\fn+\fr}$}
\multiput(158,13.5)(8,0){8}{\line(1,0){4}}
\put(156,5){$\underbrace{\hspace{65pt}}_{\fr}$}
\put(255,50){\line(-1,0){35}}
\put(255,50){\line(-2,1){35}}
\put(255,50){\line(-2,-1){35}}
\put(255,50){\line(-1,1){35}}
\put(255,50){\line(-1,-1){35}}
\put(235,50){\vector(1,0){5}}
\put(235,40){\vector(2,1){10}}
\put(235,60){\vector(2,-1){10}}
\put(235,30){\vector(1,1){10}}
\put(235,70){\vector(1,-1){10}}
\put(245,70){$\Sigma(p)$}
\put(255,50){\circle*{7}}
\put(255,50){\vector(1,0){35}}
\put(255,50){\vector(2,1){35}}
\put(255,50){\vector(2,-1){35}}
\put(255,50){\vector(1,1){35}}
\put(255,50){\vector(1,-1){35}}
\put(295,48){$\left.
\begin{array}{c} \fa_{\fn+\fr+1} \\[1.2pt] \fa_{\fn+\fr+2} \\[1.2pt] \fa_{\fn+\fr+3} \\[1.2pt] \vdots 
\\[1.2pt] \fa_{\fn+\fr+\fm} 
\end{array}\right\}\ \fm$}
\end{picture}
\end{center}
\caption{General gluing of two vertices\label{fig:gen2V}}
\end{figure}
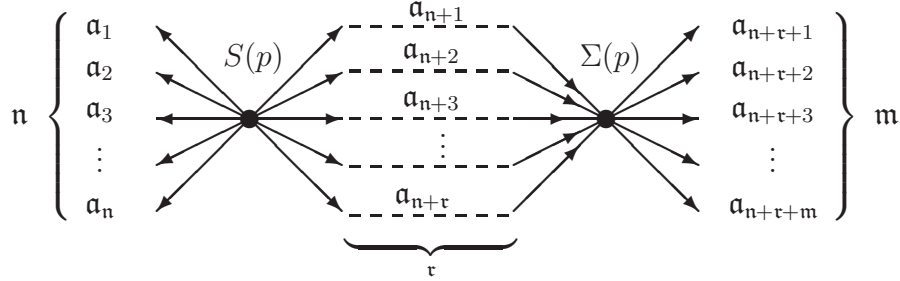

We now turn to the case of two vertices linked by $\fr$ lines, as shown 
in figure \ref{fig:gen2V}.

Following the same technics as in section \ref{sec:sim2V}, it is 
clear that the construction of the quantum field on the graph is 
equivalent to the determination of the total scattering matrix for 
the complete graph.
\subsection{General case\label{sec:glcal}}
 As in the 
previous case, we divide the local $S$ matrices according to the $\fr$ 
lines that are common to the two vertices:
\begin{equation}
S(p)=\left(\begin{array}{cc}
S_{11}(p)  & S_{12}(p) \\[1.2ex]
S_{21}(p) & S_{22}(p) 
\end{array}\right)
\mb{;}
\Sigma(p)=\left(\begin{array}{cc}
\Sigma_{11}(p)  &   \Sigma_{12}(p) \\[1.2ex]
\Sigma_{21}(p) & \Sigma_{22}(p)
\end{array}\right)
\end{equation}
where the block submatrices have sizes $\fn\times \fn$, $\fn\times \fr$, 
$\fr\times \fn$, $\fr\times \fr$ in $S(p)$:
\begin{equation}
\begin{array}{ll}
S_{11}(p)=\left(\begin{array}{ccc}
s_{11}(p) & \ldots & s_{1\fn}(p) \\
\vdots & & \vdots \\
s_{\fn1}(p) & \ldots & s_{\fn\fn}(p)
\end{array}\right)\qquad
&
S_{12}(p)=\left(\begin{array}{ccc}
s_{1,\fn+1}(p) & \ldots & s_{1,\fn+\fr}(p) \\
\vdots & & \vdots \\
s_{\fn,\fn+1}(p) & \ldots & s_{\fn,\fn+\fr}(p)
\end{array}\right)
\\[4.6ex]
S_{21}(p)=\left(\begin{array}{ccc}
s_{\fn+1,1}(p) & \ldots & s_{\fn+1,\fn}(p) \\
\vdots & & \vdots \\
s_{\fn+\fr,1}(p) & \ldots & s_{\fn+\fr,\fn}(p)
\end{array}\right)
&
S_{22}(p)=\left(\begin{array}{ccc}
s_{\fn+1,\fn+1}(p) & \ldots & s_{\fn+1,\fn+\fr}(p) \\
\vdots & & \vdots \\
s_{\fn+\fr,\fn+1}(p) & \ldots & s_{\fn+\fr,\fn+\fr}(p)
\end{array}\right)
\end{array}
\end{equation}
 and sizes $\fm\times \fm$, $\fm\times \fr$, 
$\fr\times \fm$, $\fr\times \fr$ in $\Sigma(p)$:
\begin{equation}
\begin{array}{ll}
\Sigma_{11}(p)=\left(\begin{array}{ccc}
\sigma_{11}(p) & \ldots & \sigma_{1\fr}(p) \\
\vdots & & \vdots \\
\sigma_{\fr1}(p) & \ldots & \sigma_{\fr\fr}(p)
\end{array}\right)
& 
\Sigma_{12}(p)=\left(\begin{array}{ccc}
\sigma_{1,\fr+1}(p) & \ldots & \sigma_{1,\fm+\fr}(p) \\
\vdots & & \vdots \\
\sigma_{\fr,\fr+1}(p) & \ldots & \sigma_{r,\fm+\fr}(p)
\end{array}\right)\qquad
\\[4.6ex]
\Sigma_{21}(p)=\left(\begin{array}{ccc}
\sigma_{\fr+1,1}(p) & \ldots & \sigma_{\fr+1,\fr}(p) \\
\vdots & & \vdots \\
\sigma_{\fm+\fr,1}(p) & \ldots & \sigma_{\fm+\fr,\fr}(p)
\end{array}\right)
& 
\Sigma_{22}(p)=\left(\begin{array}{ccc}
\sigma_{\fr+1,\fr+1}(p) & \ldots & \sigma_{\fr+1,\fm+\fr}(p) \\
\vdots & & \vdots \\
\sigma_{\fm+\fr,\fr+1}(p) & \ldots & \sigma_{\fm+\fr,\fm+\fr}(p)
\end{array}\right)
\end{array}
\end{equation}
The modes $\wh A(p)$ and $\wh B(p)$ on each local vertex are 
decomposed accordingly:
\begin{equation}
\wh A(p)=\left(\begin{array}{c} 
A_{1}(p) \\[1.2ex] A_{2}(p)
\end{array}\right)
\mb{;}
\wh B(p)=\left(\begin{array}{c} 
A_{2}(-p) \\[1.2ex] A_{3}(p)
\end{array}\right)
\end{equation}
where
\begin{equation}
 A_{1}(p)=\left(\begin{array}{c} 
\fa_{1}(p) \\  \vdots \\ \fa_{\fn}(p)
\end{array}\right)
\mb{;}
 A_{2}(p)=\left(\begin{array}{c} 
 \fa_{\fn+1}(p) \\ \vdots \\ \fa_{\fn+\fr}(p)
\end{array}\right)
\mb{;}
 A_{3}(p)=\left(\begin{array}{c} 
 \fa_{\fn+\fr+1}(p) \\ \vdots \\ \fa_{\fn+\fm+\fr}(p)
\end{array}\right)\,.
\end{equation}

The calculation follows the same lines as in section 
\ref{sec:sim2V} 
and we get
\begin{equation}
\cA(p)=S_{tot}(p)\,\cA(-p) \mb{with} \cA(p)=\left(\begin{array}{c} 
A_{1}(p) \\[1.2ex] A_{3}(p) \end{array}\right)
\label{eq:boundStot}
\end{equation}
and
\begin{equation}
S_{tot}(p) = \left(\begin{array}{cc}
\displaystyle
S_{11}(p)
+S_{12}(p)\,D(p)^{-1}\,\Sigma_{11}(p)\,S_{21}(p)
& \displaystyle
S_{12}(p)\,D(p)^{-1}\,\Sigma_{12}(p)
\\[2.1ex]
\displaystyle
\Sigma_{21}(p)\,\wt D(p)^{-1}\,S_{21}(p)
&\displaystyle
\Sigma_{22}(p)
+\Sigma_{21}(p)\,\wt D(p)^{-1}\,S_{22}(p)\,\Sigma_{12}(p)
\end{array}\right)
\label{eq:Stot-gen2V}
\end{equation}
where 
\begin{equation}
D(p)=\II_{\fr}-\Sigma_{11}(p)\,S_{22}(p)
\mb{and} \wt D(p)=\II_{\fr}-S_{22}(p)\,\Sigma_{11}(p)
\end{equation}
 is now an $\fr\times \fr$ matrix supposed to 
be invertible (which is true for generic values of $d$, the distance between the 
two vertices). 

One checks easily that the formulas 
(\ref{eq:Stot-gen2V}) are identical to the one given by the 
star-product approach, see e.g. formula (33) in \cite{Schra2} and formula (3.4) 
in \cite{Schra}. The matrices $D(p)^{-1}$ and $\wt D(p)^{-1}$ in the 
present paper correspond (through a series expansion) 
to the matrices $K_{1}$ and $K_{2}$ there, and the assumption of 
invertibility of $D(p)$ and $\wt D(p)$ is the compatibility condition 
assumed in \cite{Schra2,Schra}. In the language of \cite{Schra2,Schra}, we have made the 
generalized star-product $S(p)\,*_{W(p)}\,\Sigma(p)$. 

The present approach allows also to reconstruct
the modes in between the two vertices from the ones outside as:
\begin{equation}
A_{2}(p) = \wt D(p)^{-1}\,\Big( S_{21}(p)\, A_{1}(-p)+ S_{22}(p)\,\Sigma_{12}(p)\, 
A_{3}(-p)\Big)\,.
\label{eq:inner-gen}
\end{equation}
As in section \ref{sec:sim2V}, there is an additional consistency 
relation that is automatically satisfied if $S(p)$ and $\Sigma(p)$ obey 
the consistency relation (\ref{cons}). The proof is given in appendix 
\ref{app:cons}.
 In this case, $S_{tot}(p)$ also obeys this relation. The same si true 
 for the unitarity relation.
When one takes $\fr=1$, we recover the case of section \ref{sec:sim2V}.

\subsection{Case of identical vertices}
To simplify the expression of $S_{tot}(p)$ given above, we now focus 
on identical vertices. As already mentioned, due to the different labeling 
of the edges, the orientation of the edges and the choice of the origin on the connecting edge, 
the local scattering matrices are not identical, but rather obey
\begin{equation}
S(p) = \left(\begin{array}{cc} 
S_{11}(p) & S_{12}(p) \\ 
S_{21}(p) & S_{22}(p)
\end{array}\right)
\mb{and}
\Sigma(p) = \left(\begin{array}{cc} 
e^{2ipd}\,S_{22}(-p) & e^{ipd}\,S_{21}(-p) \\ 
e^{ipd}\,S_{12}(-p) & S_{11}(-p)
\end{array}\right)
\end{equation}
where we assumed that the distance between the two vertices is $d$, 
whatever the connecting edge on which it is measured. Since the 
vertices are identical, one has $\fn=\fm$, and $S_{11}(p)$ is a 
$\fn\times\fn$ matrix, while $S_{22}(p)$ is $\fr\times\fr$.

Then, using consistency relations (\ref{cons1})-(\ref{cons2}), which in particular implies
\begin{equation}
S_{12}(p)\,\Big(\II_{\fn}-e^{2ipd}\,S_{22}(-p)\,S_{22}(p)\Big)=\Big(\II_{\fn}-e^{2ipd}\,S_{11}(p)\,S_{11}(-p)\Big)\,
S_{12}(p)\,,
\end{equation}
one can rewrite (\ref{eq:Stot-gen2V}) as
\begin{equation}
S_{tot}(p) = \left(\begin{array}{cc}
\displaystyle
(1-e^{2ipd})\,D_{1}(p)^{-1}\,S_{11}(p)
& \displaystyle
e^{ipd}\,D_{1}(p)^{-1}\,\Big(\II_{\fr}-\,S_{11}(p)\,S_{11}(-p)\Big)
\\[2.1ex]
\displaystyle
e^{ipd}\,\wt D_{1}(p)^{-1}\,\Big(\II_{\fr}-\,S_{11}(-p)\,S_{11}(p)\Big)
&\displaystyle
(1-e^{2ipd})\,\wt D_{1}(p)^{-1}\,S_{11}(-p)
\end{array}\right)
\label{eq:Stot-gen2V-sim}
\end{equation}
where 
\begin{equation}
D_{1}(p)=\II_{\fr}-e^{2ipd}\,S_{11}(p)\,S_{11}(-p)
\mb{and}
\wt D_{1}(p)=\II_{\fr}-e^{2ipd}\,S_{11}(-p)\,S_{11}(p)\,.
\end{equation}
Remark that the total scattering matrix is built on the block 
submatrix $S_{11}(p)$ solely. 

The modes on the inner edges are given by:
\begin{equation}
A_{2}(p) = \Big(\II_{\fn}-e^{2ipd}\,S_{22}(-p)\,S_{22}(p)\Big)^{-1}\,
\Big( S_{21}(p)\, A_{1}(-p)+ e^{ipd}\,S_{22}(p)\,S_{21}(-p)\, 
A_{3}(-p)\Big)\,.
\end{equation}

\subsection{Approximation for small distance\label{sec:shortexp}}
Taking the limit $d\to0$, one gets the trivial scattering matrix
\begin{eqnarray}
S_{tot}(p)\Big\vert_{d=0} &=& \left(\begin{array}{cc}
\displaystyle
0 & \II_{\fn}\\
\displaystyle
\II_{\fn} & 0
\end{array}\right)
\end{eqnarray}
corresponding to $\fn$ non-connected infinite lines.
%  as depicted in figure \ref{fig:Vtriv}.
% 
% \begin{figure}[htb]
% \begin{center}
% \begin{picture}(200,90) \thicklines
% \put(30,35){$
% \begin{array}{c} \fa_{1} \\[1.2pt] \fa_{2} \\[1.2pt] \fa_{3} 
% \\[1.2pt] \vdots \\[1.2pt] \fa_{\fn} 
% \end{array} $}
% \put(70,72){\vector(-1,0){10}}
% \put(70,55){\vector(-1,0){10}}
% \put(70,37){\vector(-1,0){10}}
% \put(70,19){\vector(-1,0){10}}
% \put(70,0.5){\vector(-1,0){10}}
% \put(60,72){\line(1,0){80}}
% \put(60,55){\line(1,0){80}}
% \put(60,37){\line(1,0){80}}
% \put(60,19){\line(1,0){80}}
% \put(60,0.5){\line(1,0){80}}
% \put(130,72){\vector(1,0){10}}
% \put(130,55){\vector(1,0){10}}
% \put(130,37){\vector(1,0){10}}
% \put(130,19){\vector(1,0){10}}
% \put(130,0.5){\vector(1,0){10}}
% \put(150,35){$
% \begin{array}{c} \fa_{\fn+1} \\[1.2pt] \fa_{\fn+2} \\[1.2pt] \fa_{\fn+3} \\[1.2pt] \vdots 
% \\[1.2pt] \fa_{2\fn} 
% \end{array}$}
% \end{picture}
% \end{center}
% \caption{Trivial scattering vertices\label{fig:Vtriv}}
% \end{figure}

Thus, one could think of an expansion of $S_{tot}(p)$ in terms 
of the distance $d$ to get new scattering matrices. 
However, the physical data, such as the conductance (see 
section \ref{sec:conduc}), rely heavily on the 
pole structure of the scattering matrix. Hence, before we perform an 
approximation of (\ref{eq:Stot-gen2V-sim}) for small $d$, we rewrite 
it as (for $\ft=\tan(pd/2)$): 
\begin{eqnarray}
S_{tot}(p) &=& \left(\begin{array}{cc}
\displaystyle
S_{11}'(p) & S_{12}'(p)\\
\displaystyle
S_{21}'(p) & S_{22}'(p)
\end{array}\right)
\\
S_{11}'(p) &=&
\displaystyle
-4i\,\ft\,\fD(p)^{-1}\,S_{11}(p)
\mb{;}
S_{12}'(p)\ =\  (1+\ft^2)\,\fD(p)^{-1}\,
\Big(\II-S_{11}(p)\,S_{11}(-p)\,\Big)
\\
\fD(p) &=& 
(1-i\,\ft)^2\,\II-(1+i\,\ft)^2\,S_{11}(p)\,S_{11}(-p)
\\
S_{21}'(p)  &=& (1+\ft^2)\,\wt\fD(p)^{-1}\,
\Big(\II-S_{11}(-p)\,S_{11}(p)\,\Big)
 \mbox{ ; }
S_{22}'(p)\ =\ -4i\,\ft\,\wt\fD(p)^{-1}\,S_{11}(-p)
\qquad\quad \\
\wt\fD(p) &=& 
(1-i\,\ft)^2\,\II+(1+i\,\ft)^2\,S_{11}(-p)\,S_{11}(p)
\,.
\end{eqnarray}
Then, the approximation is done using the expansion $\ft\sim pd/2$, but 
keeping the possible fractions entering the formulas above.
 We detail below in section \ref{sec:scalinv} this expansion for some
examples: it will show how the pole structure is (partially) preserved 
in such an expansion. In section \ref{sec:conduc}, we apply this expansion 
to the calculation of conductance. In particular, we will 
show its consistency by
comparing in one example this approximation to the full calculation of 
the conductance.

\subsection{Example 2: the loop graph \label{sec:loop}}
In this case, one considers two vertices with three edges each, two 
of them being glued together. The local $S$-matrices are $3\times 3$, 
and the total $S$-matrix (after gluing) is $2\times2$. In the 
notation of the previous section, we have $\fn=\fm=3$ and $\fr=2$.  
\begin{figure}[htb]
\begin{center}
\begin{picture}(260,50) \thicklines
\put(20,20){$ \fa_{1} $}
\put(70,20){\vector(-1,0){35}}
\put(60,30){$S(p)$}
\put(70,20){\circle*{7}}
\put(70,20){\vector(2,1){35}}
\put(70,20){\vector(2,-1){35}}
\put(128,43){$\fa_{4}$}
\multiput(108,38)(8,0){8}{\line(1,0){4}}
\put(130,06){$\fa_{2}$}
\multiput(108,02)(8,0){8}{\line(1,0){4}}
\put(205,20){\line(-2,1){35}}
\put(205,20){\line(-2,-1){35}}
\put(185,10){\vector(2,1){10}}
\put(185,30){\vector(2,-1){10}}
\put(200,30){$\Sigma(p)$}
\put(205,20){\circle*{7}}
\put(205,20){\vector(1,0){35}}
\put(245,18){$ \fa_{3}$}
\end{picture}
\end{center}
\caption{The loop\label{fig:loop}}
\end{figure}
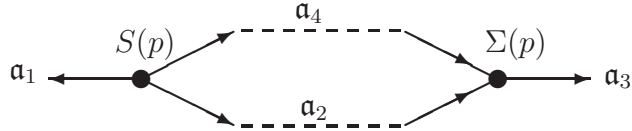

The decomposition of the local $S$ matrices reads
\begin{equation}
S(p) = \left(\begin{array}{c|cc} 
s_{11}(p) & s_{12}(p) & s_{13}(p) \\
\hline
s_{21}(p) & s_{22}(p) & s_{23}(p) \\
s_{31}(p) & s_{32}(p) & s_{33}(p) 
\end{array}\right)
\mb{and}
\Sigma(p) = \left(\begin{array}{cc|c} 
\sigma_{11}(p) & \sigma_{12}(p) & \sigma_{13}(p) \\
\sigma_{21}(p) & \sigma_{22}(p) & \sigma_{23}(p) \\
\hline
\sigma_{31}(p) & \sigma_{32}(p) & \sigma_{33}(p) 
\end{array}\right)\,.
\end{equation}
Again, the lines drawn within the matrices indicate the block 
submatrices we consider. The
 local boundary conditions have the form
\begin{equation}
\left(\begin{array}{c} \fa_{1}(p) \\ \fa_{2}(p) \\ \fa_{4}(p) 
\end{array}\right)\ =\ S(p)\, 
\left(\begin{array}{c} \fa_{1}(-p) \\ \fa_{2}(-p) \\ \fa_{4}(-p) 
\end{array}\right)
\mb{and}
\left(\begin{array}{c} \fa_{2}(-p) \\ \fa_{4}(-p) \\ \fa_{3}(p) 
\end{array}\right)\ =\ \Sigma(p)\, 
\left(\begin{array}{c} \fa_{2}(p) \\ \fa_{4}(p) \\ \fa_{3}(-p) 
\end{array}\right)\,.
\end{equation}

We will focuss on identical vertices:
\begin{equation}
\Sigma(p)=W(p)\,P^{-1}\,S(-p)\,P\,W(p)
\mb{with} P=\left(\begin{array}{ccc} 
0 & 0 & 1 \\ 1 & 0 & 0 \\ 0 & 1 & 0
\end{array}\right)\ ;\ 
 W(p)=\mbox{diag}(e^{ipd_{2}},e^{ipd_{4}},1)
\end{equation}
where $d_{a}$ is the distance between the two vertices, measured on 
the edges $a=2,4$. It leads to
\begin{equation}
\Sigma(p) = \left(\begin{array}{cc|c} 
s_{22}(-p)\,e^{2ipd_{2}} & s_{23}(-p)\,e^{ip(d_{2}+d_{4})} 
& s_{21}(-p)\,e^{ipd_{2}} 
\\
s_{32}(-p)\,e^{ip(d_{2}+d_{4})} & s_{33}(-p)\,e^{2ipd_{4}} 
& s_{31}(-p)\,e^{ipd_{4}}\\[1.ex]
\hline\rule{0ex}{2.4ex}
s_{12}(-p)\,e^{ipd_{2}} & s_{13}(-p)\,e^{ipd_{4}} & s_{11}(-p) 
\end{array}\right)\,.
\end{equation}
To get simple expressions, we suppose that 
$d_{2}=d_{4}=d/2$, where $d$ is the total length of the loop. 
The total scattering matrix takes the form (with $\ft=\tan(dp/2)$):
\begin{eqnarray}
S_{tot}(p) &=& 
\frac{1}{\cN(p)}
\,\left(\begin{array}{cc} 
-4i\ft\,s_{11}(p) 
& (1+\ft^2)\,\big(1-s_{11}(p)\,s_{11}(-p)\big)
\\
(1+\ft^2)\,\big(1-s_{11}(p)\,s_{11}(-p)\big)
& -4i\ft\,s_{11}(-p) 
\end{array}\right)
\nonumber\\[2.1ex]
\cN(p) &=& (1-i\,\ft)^2-(1+i\,\ft)^2\,s_{11}(-p)\,s_{11}(p)\,.
\label{eq:Sloop-gen}
\end{eqnarray}
It corresponds for the total graph to a boundary condition:
\begin{equation}
\left(\begin{array}{c} \fa_{1}(p) \\  \fa_{3}(p) 
\end{array}\right)\ =\ S_{tot}(p)\, 
\left(\begin{array}{c} \fa_{1}(-p) \\ \fa_{3}(-p) 
\end{array}\right)\,.
\end{equation}
The inner modes read
\begin{eqnarray}
\fa_{2}(p) &=& -\frac{s_{21}(p)}{\cN(p)}\,\Big( \fa_{1}(-p)
+e^{ipd}\,s_{11}(-p)\,\fa_{3}(-p)\Big)
\\
\fa_{4}(p) &=& -\frac{s_{31}(p)}{\cN(p)}\,\Big( \fa_{1}(-p)+
e^{ipd}\,s_{11}(-p)\,\fa_{3}(-p)\Big)\,.
\end{eqnarray}

\subsubsection{Expansion for short distances}
The general formula (\ref{eq:Sloop-gen}) simplifies to
\begin{eqnarray}
S_{tot}(p) &\sim& 
\frac{1}{\cN_{0}(p)}
\,\left(\begin{array}{cc} 
-2ipd\,s_{11}(p) & 1-s_{11}(p)\,s_{11}(-p)
\\
1-s_{11}(p)\,s_{11}(-p) & -2ipd\,s_{11}(-p) 
\end{array}\right)
\nonumber\\[2.1ex]
\cN_{0}(p) &=& \left(1-i\,\frac{dp}{2}\right)^2
-\left(1+i\,\frac{dp}{2}\right)^2\,s_{11}(-p)\,s_{11}(p)
\end{eqnarray}
where now $s_{11}(p)$ is a scalar function. If one assumes 
furthermore that $s_{11}(p)$ is a constant (see section 
\ref{sec:scalinv}), 
the expansion leads to a total scattering matrix with two simple 
poles $\frac{2i}{d}\,\frac{s_{11}\pm1}{s_{11}\mp1}$. 

\subsection{Example 3: the tadpole graph\label{sec:tad}}
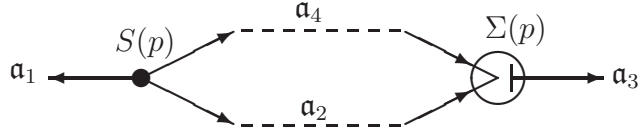
\begin{figure}[htb]
\begin{center}
\begin{picture}(260,50) \thicklines
\put(20,20){$ \fa_{1} $}
\put(70,20){\vector(-1,0){35}}
\put(60,30){$S(p)$}
\put(70,20){\circle*{7}}
\put(70,20){\vector(2,1){35}}
\put(70,20){\vector(2,-1){35}}
\put(128,43){$\fa_{4}$}
\multiput(108,38)(8,0){8}{\line(1,0){4}}
\put(130,06){$\fa_{2}$}
\multiput(108,02)(8,0){8}{\line(1,0){4}}
\put(205,20){\line(-2,1){35}}
\put(205,20){\line(-2,-1){35}}
\put(185,10){\vector(2,1){10}}
\put(185,30){\vector(2,-1){10}}
\put(200,35){$\Sigma(p)$}
\put(205,20){\circle{20}}
\put(210,20){\vector(1,0){35}}
\put(210,15){\line(0,1){10}}
\put(248,18){$\fa_{3}$}
\end{picture}
\end{center}
\caption{The tadpole (plus a half-line)\label{fig:tadpol}}
\end{figure}

The tadpole is constructed as a special case of loop, where one of 
the vertex is fully transmitting between two edges, and purely 
reflexive in the third edge (with coefficient 1). 
In this way, we get a system with a 
tadpole and a decoupled half-line, as depicted in figure \ref{fig:tadpol}.
The local scattering matrices of this graph are not identical: 
the first one has the general form
\begin{equation}
S(p) = \left(\begin{array}{c|cc} 
s_{11}(p) & s_{12}(p) & s_{13}(p) \\
\hline
s_{21}(p) & s_{22}(p) & s_{23}(p) \\
s_{31}(p) & s_{32}(p) & s_{33}(p) 
\end{array}\right)
= \left(\begin{array}{cc} S_{11}(p) & S_{12}(p) \\
S_{21}(p) & S_{22}(p) \end{array}\right)
\end{equation}
while the one associated to the purely reflexive vertex reads
\begin{equation}
\Sigma(p) = \left(\begin{array}{cc|c} 
0 & e^{ip(d_{2}+d_{4})} & 0 \\ e^{ip(d_{2}+d_{4})} & 0 & 0 \\ 
\hline 0 & 0 & 
1 \end{array}\right)
= \left(\begin{array}{cc} \Sigma_{11}(p) & \Sigma_{12}(p) \\
\Sigma_{21}(p) & \Sigma_{22}(p) \end{array}\right)
\label{eq:sig-tadpol}
\end{equation}
where $d_{2}$ and $d_{4}$ are the distances between the two vertices, 
measured on the edges 2 and 4 respectively, so that the length of the 
loop in the tadpole is $\ell=d_{2}+d_{4}$.

From formulas (\ref{eq:Stot-gen2V})-(\ref{eq:inner-gen}), we get
\ben
S_{tot}(p) &=& \left(\begin{array}{cc}
R(p) & 0 \\ 0 & 1\end{array}\right)
\label{eq:Stot-tadpole}
\\
R(p) &=& s_{11}(p) +\frac{e^{ip\ell}}{N(p)}
\Big(R_{0}(p)+e^{ip\ell}\,R_{1}(p)\Big)
\nonu
R_{0}(p) &=& s_{12}(p)\,s_{31}(p) +s_{13}(p)\,s_{21}(p)  
\nonu
R_{1}(p) &=& s_{12}(p)\Big(s_{33}(p)\,s_{21}(p)
-s_{23}(p)\,s_{31}(p)\Big)+
s_{13}(p)\Big(s_{22}(p)\,s_{31}(p)-s_{32}(p)\,s_{21}(p)\Big)
\nonu
N(p) &=& 
\big(1-e^{ip\ell}\,s_{23}(p)\big)\big(1-e^{ip\ell}\,s_{32}(p)\big)
 - e^{2ip\ell}\,s_{22}(p)s_{33}(p)
\nonumber
\een
 The modes in between the two vertices are 
reconstructed from the ones outside as:
\ben
\left(\begin{array}{c} \fa_{2}(p) \\ \fa_{4}(p) \end{array}\right)
&=& \wt D(p)^{-1}\,S_{21}(p)\, \fa_{1}(-p) \nonu
&=& \left(\begin{array}{c}
s_{21}(p) +\big(s_{22}(p)s_{31}(p)-s_{32}(p)s_{21}(p)\big)\,e^{ip\ell}\\
s_{31}(p) +\big(s_{33}(p)s_{21}(p)-s_{23}(p)s_{31}(p)\big)\,e^{ip\ell}
\end{array}\right)\,\frac{\fa_{1}(-p)}{N(p)}\,.
\label{eq:inner-tadpole}
\een
As expected, the mode $\fa_{3}(p)$ on the purely reflexive half-line 
decouples, and the mode(s) on the loop of the tadpole depends solely 
on $\fa_{1}(p)$, the mode on the outer line of the tadpole. This mode 
obeys a reflection boundary condition.

Again, one can perform an approximation for small distances: we will 
present it in section \ref{sec:scalinv} on particular examples that 
apply to the calculation of conductance on graphs.

\section{General gluing of more than two vertices \label{sec:genVgen}}
The construction is done by recursion: one first glues two vertices 
using the results of the previous section to get an effective vertex 
corresponding to this gluing. Then, one glues this effective vertex 
to a third one. The result is the gluing of the three vertices, that 
we can glue to a fourth one, and so on\ldots The quantum field 
follows the same rule, and, at the end, we get the field for the total 
graph in term of the generators $\{\fa_{a}(p),\,\fa_{a}^\dag(p)\}$ of 
the \textit{external} edges $a$ solely. It obeys relations (\ref{eq:cano}) 
on external edges. 

The total scattering matrix of a general graph is thus obtained 
through a recursive use of the formulas of section \ref{sec:gen2V}. 
If we denote by $S^{[j]}(p)$, $j=1,\ldots,\cn+1$, the local $S$-matrices 
of the $\cn+1$ vertices under consideration, and by $S^{[j\ldots 
k]}(p)$, $1\leq j<k\leq \cn+1$,
the $S$-matrix resulting from the gluing of the vertices $j$ to $k$, 
we get the recursion formula:
\begin{eqnarray}
S^{[1\ldots \cn+1]}_{tot}(p) &=& \left(\begin{array}{cc}
\displaystyle S^{[1\ldots \cn+1]}_{tot}(p)_{11}
& \displaystyle S^{[1\ldots \cn+1]}_{tot}(p)_{12}
\\[2.1ex]
\displaystyle S^{[1\ldots \cn+1]}_{tot}(p)_{21}
&\displaystyle S^{[1\ldots \cn+1]}_{tot}(p)_{22}
\end{array}\right) \\[2.1ex]
S^{[1\ldots \cn+1]}_{tot}(p)_{11} &=& 
\displaystyle
\wt S^{[1\ldots \cn]}_{11}(p)
+\wt S^{[1\ldots \cn]}_{12}(p)\,D(p)^{-1}\, S^{[\cn+1]}_{11}(p)
\,\wt S^{[1\ldots \cn]}_{21}(p)
\\[1.2ex]
S^{[1\ldots \cn+1]}_{tot}(p)_{12} &=& 
 \displaystyle
\wt S^{[1\ldots \cn]}_{12}(p)\,D(p)^{-1}\,S^{[\cn+1]}_{12}(p)
\\[1.2ex]
S^{[1\ldots \cn+1]}_{tot}(p)_{21} &=& 
\displaystyle
S^{[\cn+1]}_{21}(p)\,\wt D(p)^{-1}\,\wt S^{[1\ldots \cn]}_{21}(p)
\\[1.2ex]
S^{[1\ldots \cn+1]}_{tot}(p)_{22} &=& 
\displaystyle
S^{[\cn+1]}_{22}(p)
+S^{[\cn+1]}_{21}(p)\,\wt D(p)^{-1}\,\wt S^{[1\ldots \cn]}_{22}(p)
\,S^{[\cn+1]}_{12}(p)
\\[1.2ex]
D(p) &=& \II-S^{[\cn+1]}_{11}(p)\,\wt S^{[1\ldots \cn]}_{22}(p)
\mb{;}
\wt D(p) \ =\ \II-\wt S^{[1\ldots \cn]}_{22}(p)\,S^{[\cn+1]}_{11}(p)
\end{eqnarray}
where $D(p)$ and $\wt D(p)$ are supposed to be invertible. 
$\wt S^{[1\ldots \cn]}(p)$ is 
 deduced from the scattering matrix $S^{[1\ldots \cn]}(p)$ obtained from 
the previous step through a reordering of the rows and columns such 
that the modes `glued' in the step appear at the right place (see 
section \ref{sec:star-triang}).
Of course, the decomposition of the $S$-matrices into 
submatrices
$S_{11}$, $S_{12}$, $S_{21}$ and $S_{22}$ and the size of these 
submatrices depend on the number of edges that are glued between each 
vertices. 

\subsection{Example 4: star-triangle relation \label{sec:star-triang}}
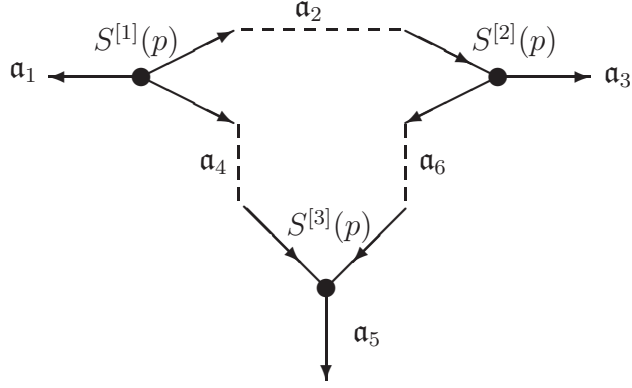
\begin{figure}[htb]
\begin{center}
\begin{picture}(260,130) \thicklines
\put(20,100){$ \fa_{1} $}
\put(70,100){\vector(-1,0){35}}
\put(52,110){$S^{[1]}(p)$}
\put(70,100){\circle*{7}}
\put(70,100){\vector(2,1){35}}
\put(70,100){\vector(2,-1){35}}
\put(128,123){$\fa_{2}$}
\multiput(108,118)(8,0){8}{\line(1,0){4}}
\put(92,65){$\fa_{4}$}
\multiput(107,82)(0,-8){4}{\line(0,-1){5}}
\put(175,65){$\fa_{6}$}
\multiput(170,82)(0,-8){4}{\line(0,-1){5}}
\put(140,20){\line(1,1){30}}
\put(120,40){\vector(1,-1){10}}
\put(140,20){\line(-1,1){31}}
\put(125,40){$S^{[3]}(p)$}
\put(140,20){\circle*{7}}
\put(140,20){\vector(0,-1){35}}
\put(150,0){$ \fa_{5}$}
\put(205,100){\line(-2,1){35}}
\put(205,100){\vector(-2,-1){35}}
 \put(160,40){\vector(-1,-1){10}}
\put(185,110){\vector(2,-1){10}}
\put(195,110){$S^{[2]}(p)$}
\put(205,100){\circle*{7}}
\put(205,100){\vector(1,0){35}}
\put(245,98){$ \fa_{3}$}
\end{picture}
\end{center}
\caption{The triangle\label{fig:triangle}}
\end{figure}

We consider a graph constituted with $\cn=3$ identical vertices possessing three edges each, coupled as in 
figure \ref{fig:triangle}.
The local boundary conditions are given by
\begin{eqnarray}
\left(\begin{array}{c} \fa_{1}(p) \\ \fa_{4}(p) \\ \fa_{2}(p) 
\end{array}\right)\ =\ S^{[1]}(p)\, 
\left(\begin{array}{c} \fa_{1}(-p) \\ \fa_{4}(-p) \\ \fa_{2}(-p) 
\end{array}\right)
&\mb{;}&
\left(\begin{array}{c} \fa_{2}(-p) \\ \fa_{3}(p) \\ \fa_{6}(p) 
\end{array}\right)\ =\ S^{[2]}(p)\, 
\left(\begin{array}{c} \fa_{2}(p) \\ \fa_{3}(-p) \\ \fa_{6}(-p) 
\end{array}\right)
\nonumber\\[1.2ex]
\left(\begin{array}{c} \fa_{4}(-p) \\ \fa_{6}(-p) \\ \fa_{5}(p) 
\end{array}\right)\ =\ S^{[3]}(p)\, 
\left(\begin{array}{c} \fa_{4}(p) \\ \fa_{6}(p) \\ \fa_{5}(-p) 
\end{array}\right)\,.&&
\label{eq:boundS3}
\end{eqnarray}

We first construct $S^{[12]}(p)$ as the gluing of $S^{[1]}(p)$ and 
$S^{[2]}(p)$. Since the vertices are chosen identical, we have
\beq
S^{[2]}(p) = W_{2}(p)\,P_{2}\,S^{[1]}(-p)\,P_{2}^{-1}\,W_{2}(p)
= \left(\begin{array}{c|cc} 
s_{33}(-p)\,e^{2ipd} & s_{31}(-p)\,e^{ipd} & s_{32}(-p)\,e^{ipd} 
\\[1.ex]
\hline\rule{0ex}{2.4ex}
s_{13}(-p)\,e^{ipd} & s_{11}(-p) & s_{12}(-p) \\
s_{23}(-p)\,e^{ipd} & s_{21}(-p) & s_{22}(-p) 
\end{array}\right)\,.
\end{equation}

During this first gluing, the mode $\fa_{2}(p)$ is the only inner 
mode. The modes $\fa_{4}(p)$ and $\fa_{6}(p)$, which are inner modes 
of the full graph, are considered as outer modes for a while.
We can apply the results of section \ref{sec:tree} for the tree graph 
to get $S^{[12]}(p)$: it is in fact of the form
(\ref{eq:Stot-gen2V-sim}), where $S_{11}(p)$ is the $2\times 2$ 
upper left submatrix of $S^{[1]}(p)$.

The inner mode $\fa_{2}(p)$ is constructed from the `outer' modes 
$\fa_{a}(p)$, $a=1,3,4,6$:
\begin{eqnarray}
\fa_{2}(p) \ =\ \frac{1}{1-e^{2ipd}\,s_{33}(p)\,s_{33}(-p)} &\Big\{&
\! s_{31}(p)\,\fa_{1}(-p)+s_{32}(p)\,\fa_{4}(-p)
\label{eq:tri-a2}\\
&+&\left.
e^{ipd}\,s_{33}(p)\Big(s_{31}(-p)\,\fa_{3}(-p)+s_{32}(-p)\,\fa_{6}(-p)
\Big)\ \right\}\,.\nonumber
\end{eqnarray}
These `outer' modes obey the boundary condition
\begin{equation}
\left(\begin{array}{c} \fa_{1}(p) \\ \fa_{4}(p)\\ \fa_{3}(p) \\ \fa_{6}(p)
\end{array}\right) = S^{[12]}(p)\,
\left(\begin{array}{c} \fa_{1}(-p) \\ \fa_{4}(-p)\\ \fa_{3}(-p) \\ \fa_{6}(-p)
\end{array}\right)\,.
\label{eq:boundS12}
\end{equation}

We now turn to the second stage of the gluing: we glue $S^{[12]}(p)$ 
with $S^{[3]}(p)$. Sticking to the identical vertices case, 
we take $S^{[3]}(p)$ to be
\begin{equation}
S^{[3]}(p)=W_{3}(p)\,P_{3}\,S^{[1]}(p)\,P_{3}^{-1}\,W_{3}(p)
=\left(\begin{array}{cc|c}
s_{11}(p)\,e^{2ipd} & s_{13}(p)\,e^{2ipd} & s_{12}(p)\,e^{ipd} 
\\
s_{31}(p)\,e^{2ipd} & s_{33}(p)\,e^{2ipd} & s_{32}(p)\,e^{ipd}  \\
\hline
s_{21}(p)\,e^{ipd} & s_{23}(p)\,e^{ipd} & s_{22}(p) 
\end{array}\right)\,,
\end{equation}
so that we have a local boundary condition as in (\ref{eq:boundS3}).
We have chosen the distance $d$ to be the same on each edge, but 
clearly nothing changes in the construction if the distance $d_{12}$ between 
the vertices 1 and 2 (appearing at the first stage) is different from the 
distances $d_{13}$ and $d_{23}$ (appearing at the second stage). 
To do the gluing, we have also to reformulate the boundary condition 
(\ref{eq:boundS12}) in the following way:
\begin{equation}
\left(\begin{array}{c} \fa_{1}(p) \\ \fa_{3}(p)\\ \fa_{4}(p) \\ \fa_{6}(p)
\end{array}\right) = \wt S^{[12]}(p)\,
\left(\begin{array}{c} \fa_{1}(-p) \\ \fa_{3}(-p)\\ \fa_{4}(-p) \\ \fa_{6}(-p)
\end{array}\right)\,.
\end{equation}
 The new  $\wt S^{[12]}(p)$ is deduced from the original $ S^{[12]}(p)$ through the 
reordering:
\begin{equation}
\wt S^{[12]}(p) = P_{12}\, S^{[12]}(p)\,P_{12}^{-1}
\mb{with} P_{12}=\left(\begin{array}{cccc}
1 & 0 & 0 & 0 \\ 0 & 0 & 1 & 0 \\ 0 & 1 & 0 & 0 \\ 0 & 0 & 0 & 1 
\end{array}\right)
\end{equation} 
Then, one just uses formulas (\ref{eq:Stot-gen2V}) with $\wt S^{[12]}(p)$ playing 
the role of $S(p)$ and $S^{[3]}(p)$ the role of $\Sigma(p)$. In this 
way, we get a scattering matrix $S_{tot}(p)$ for a `global' vertex with 
three edges, equivalent from `outside' to the three original vertices: 
this is the star-triangle relation. Indeed, the boundary condition for outer 
modes reads
\begin{equation}
\left(\begin{array}{c} \fa_{1}(p) \\ \fa_{3}(p) \\ \fa_{5}(p)
\end{array}\right) = S_{tot}(p)\,
\left(\begin{array}{c} \fa_{1}(-p) \\ \fa_{3}(-p) \\ \fa_{5}(-p)
\end{array}\right)
\end{equation}
and the inner modes $\fa_{4}(p)$ and $\fa_{6}(p)$ are obtained 
through relation (\ref{eq:inner-gen}) with 
\begin{equation}
A_{2}(p)=\left(\begin{array}{c} \fa_{4}(p) \\ \fa_{6}(p) 
\end{array}\right) \mb{;}
A_{1}(p)=\left(\begin{array}{c} \fa_{1}(p) \\ \fa_{3}(p) 
\end{array}\right) \mb{;}
A_{3}(p)= \fa_{5}(p) \,.
\end{equation}
The complete expression for the inner mode $\fa_{2}(p)$ is obtained 
using (\ref{eq:tri-a2}) and the expressions 
for $\fa_{4}(p)$ and $\fa_{6}(p)$. 
However,
the general formulas being rather complicated, we prefer not to write 
them explicitly. 
A complete example of triangle scattering matrix is given in section 
\ref{sec:triang-scal} for special (constant) local scattering matrices.

\section{Scale invariant matrices and Kirchhoff's rule\label{sec:scalinv}}
We focus on the case of identical vertices and apply the formalism to 
scale invariant matrices. Since we will deal with the examples 
treated in previous sections, that are constructed from local 
$3\times 3$ scattering matrices, we focus on scale invariant
matrices of this size. They have the form
\begin{eqnarray}
S_{\alpha} 
% &=& \II-\frac{2}{1+\alpha_1^2+\alpha_2^2}\,\left(
% \begin{array}{ccc}
%  \alpha_1^2 &  \alpha_1 \alpha_2 &  \alpha_1 \\
%   \alpha_1 \alpha_2 & \alpha_2^2 &  \alpha_2 \\
%   \alpha_1 &  \alpha_2 & 1
% \end{array}
% \right)\nonu
&=& \frac{1}{1+\alpha_1^2+\alpha_2^2}\,\left(
\begin{array}{ccc}
1-\alpha_1^2+\alpha_2^2 &  -2\alpha_1 \alpha_2 &  -2\alpha_1 \\
-2\alpha_1 \alpha_2 & 1+\alpha_1^2-\alpha_2^2 &  -2\alpha_2 \\
 -2\alpha_1 &  -2\alpha_2 & \alpha_1^2+\alpha_2^2-1
\end{array}
\right)\,.
\label{eq:Sscal}
\end{eqnarray}
We will see in the next section how to deduce the conductance 
$G_{ab}$ between edges $a$ and $b$ of a 
quantum wire from the scattering matrix of this wire. For a star 
graph and a Luttinger liquid model, the 
 conductance obeys Kirchhoff's rule
\beq
\sum_{a=1}^\fn G_{ab}=0\,,
\eeq
if the scattering matrix obeys \cite{conduc}
\beq
\sum_{a=1}^\fn S_{ab}=1\,.
\eeq
We will loosely call this relation Kirchhoff's rule (for scattering matrices).
For a general quantum wire, one may impose the Kirchhoff's rule for 
scattering matrices
\textit{locally}, i.e. on each vertex of the wire, or 
\textit{globally}, i.e. on the total scattering matrix. 
To get a scale invariant matrix (\ref{eq:Sscal}) obeying Kirchhoff's
rule, one needs to  impose the constraint
$1+\alpha_1+\alpha_2=0$.

As already mentioned, we now apply the results obtained for the examples 
1, 2, 3 and 4 
treated in previous sections to cases where the local scattering 
matrices have the form (\ref{eq:Sscal}).

\subsection{Example 1: the tree graph}
For the tree graph (see section \ref{sec:tree}) built on local scale 
invariant matrices, 
the total scattering matrix (\ref{eq:tree-gen}) takes the symmetric form:
\begin{eqnarray}
S_{tot}(p) &=& \frac{1}{\cN(p)}\,\left(\begin{array}{cc} 
S_{11}(p) & S_{12}(p) \\
S_{12}(p) & S_{11}(p) \end{array}\right)
\\[1.2ex]
\cN(p) &=& (1+\alpha_{1}^2+\alpha_{2}^2)^2 
-(1-\alpha_{1}^2-\alpha_{2}^2)^2\,e^{2ipd}\\
S_{11}(p) &=& (1+\alpha_{1}^2+\alpha_{2}^2)\,\left(\begin{array}{cc} 
 {1-\alpha_{1}^2+\alpha_{2}^2} & {-2\,\alpha_{1}\,\alpha_{2}}\\[1.2ex]
 {-2\,\alpha_{1}\,\alpha_{2}} & 
 {1+\alpha_{1}^2-\alpha_{2}^2} \end{array}\right)
\nonumber\\[1.2ex]
&&+e^{2ipd}\ (\alpha_{1}^2+\alpha_{2}^2-1) \,
\left(\begin{array}{cc} 
 {1+\alpha_{1}^2-\alpha_{2}^2} 
&  {2\,\alpha_{1}\,\alpha_{2}} \\[1.2ex]
 {2\,\alpha_{1}\,\alpha_{2}} & 
 {1-\alpha_{1}^2+\alpha_{2}^2} \end{array}\right)
\\[1.2ex]
S_{12}(p) &=& 4\,e^{ipd}\,
\left(\begin{array}{cc} 
\alpha_{1}^2 & \alpha_{1}\,\alpha_{2} \\[1.2ex]
\alpha_{1}\,\alpha_{2} & \alpha_{2}^2 \end{array}\right)\,.
\label{tree-scal}
\end{eqnarray}
The inner mode is expressed as
\begin{eqnarray}
\fa_{3}(p) &=& \frac{-2}{\cN(p)}\,\left\{ 
(\alpha_{1}^2+\alpha_{2}^2+1)\,\Big(
\alpha_{1}\,\fa_{1}(-p)+  \alpha_{2}\,\fa_{2}(-p)\Big)
\right.\nonu
&&\qquad\quad\left.+ 
(\alpha_{1}^2+\alpha_{2}^2-1)\,e^{ipd}\,\Big(
\alpha_{1}\,\fa_{4}(-p)+ 
\alpha_{2}\,\fa_{5}(-p)\Big)\right\}\,,
\end{eqnarray} 
where, for the edges, we have used the numbering given in figure \ref{fig:tree}.

When one considers the particular case $\alpha_{1}=\alpha_{2}=\pm 1$, one recovers the scattering matrix 
computed in example (IV.4) of \cite{Schra}.

\subsubsection{Approximation for short distances}
We first rewrite the scattering matrix as
\ben
S_{tot}(p) &=& \frac{4}{N(p)}\,
\left(\begin{array}{cc}
(1-\ft^2)\,A_{0}+ i\,\ft\,A_{1} & B \\ 
B & (1-\ft^2)\,A_{0}+ i\,\ft\,A_{1} 
\end{array}\right)
\\
N(p) &=& (\mu^2-1)\,\ft^2-2i(1+\mu^2)\,\ft+1-\mu^2
\mb{with}\mu=\frac{1-\alpha_{1}^2-\alpha_{2}^2}
{1+\alpha_{1}^2+\alpha_{2}^2}\\
A_{0} &=& \frac{1}{(1+\alpha_{1}^2+\alpha_{2}^2)^2}
\left(\begin{array}{cc} 
\alpha_{2}^2 & -\alpha_{1} \alpha_{2} \\
-\alpha_{1} \alpha_{2} & \alpha_{1}^2
\end{array}\right) \\
A_{1} &=& \frac{1}{(1+\alpha_{1}^2+\alpha_{2}^2)^2}
\left(\begin{array}{cc} 
\alpha_{1}^4-\alpha_{2}^4-1 
& \alpha_{1} \alpha_{2}(\alpha_{1}^2+\alpha_{2}^2) \\
\alpha_{1} \alpha_{2}(\alpha_{1}^2+\alpha_{2}^2) 
& \alpha_{2}^4-\alpha_{1}^4-1
\end{array}\right) \\
B &=& \frac{1}{(1+\alpha_{1}^2+\alpha_{2}^2)^2}
\left(\begin{array}{cc} 
\alpha_{1}^2 & \alpha_{1} \alpha_{2} \\
\alpha_{1} \alpha_{2} & \alpha_{2}^2
\end{array}\right) 
\een
where $\ft=\tan(\frac{dp}{2})$.
Taking $d=0$, we get a new $4\times4$ scattering matrix 
\begin{equation}
S^{(0)} = \frac{4}{1-\mu^2}\,
\left(\begin{array}{cc}
A_{0} & B \\ B & A_{0} 
\end{array}\right)
 = \frac{1}{\alpha_{1}^2+\alpha_{2}^2} 
\,\left(\begin{array}{cccc} 
\alpha_{2}^2 & -\alpha_{1}\,\alpha_{2} & \alpha_{1}^2 & \alpha_{1}\,\alpha_{2}\\
-\alpha_{1}\,\alpha_{2} & \alpha_{1}^2 & \alpha_{1}\,\alpha_{2} & \alpha_{2}^2 \\ 
\alpha_{1}^2 & \alpha_{1}\,\alpha_{2} & \alpha_{2}^2 & -\alpha_{1}\,\alpha_{2}\\
\alpha_{1}\,\alpha_{2} & \alpha_{2}^2 & -\alpha_{1}\,\alpha_{2} & \alpha_{1}^2 
\end{array}\right)\,.
\end{equation}
Remark that this matrix obeys the Kirchhoff's rule, whatever the values 
of $\alpha_{1}$ and $\alpha_{2}$ are, even when the local 
scattering matrices do not. 
For\footnote{Notice that if one also imposes the 
Kirchhoff's rule on local vertices, one needs to take $\eps=+1$.} 
$\alpha_{2}=\eps\,\alpha_{1}$, with $\eps=\pm1$, we get simpler 
matrices (still obeying Kirchhoff's rule):
\begin{equation}
S^{(0)} = \frac{1}{2} 
\,\left(\begin{array}{cccc} 
1 & -\eps & 1 & \eps\\
-\eps & 1 & \eps & 1 \\ 
1 & \eps & 1 & -\eps\\
\eps & 1 & -\eps & 1 
\end{array}\right)
\mb{with} \eps = \pm1\,.
\end{equation}
These matrices can be compared with the two matrices introduced in 
\cite{Cham2} for the modelisation of a condensed matter experiment 
proposal:
\begin{equation}
S^{(0)}_{ch} = \frac{1}{2} 
\,\left(\begin{array}{cccc} 
\eps & 1 & -\eps & 1\\
1 & \eps & 1 & -\eps \\ 
-\eps & 1 & \eps & 1\\
1 & -\eps & 1 & \eps 
\end{array}\right)\mb{with} \eps=\pm1\,.
\end{equation}
Indeed, one has
\begin{equation}
S^{(0)}_{ch} = U\,S^{(0)}
\mb{where} U = \left(\begin{array}{cccc} 
0 & 0 & 0 & 1\\
0 & 0 & 1 & 0 \\ 
0 & 1 & 0 & 0\\
1 & 0 & 0 & 0 
\end{array}\right)\,.
\end{equation}
Since $U\,S^{(0)}\,U=S^{(0)}$, the unitarity relations are preserved 
by this transformation, and indeed $S^{(0)}$ and $S^{(0)}_{ch}$ are 
unitary.

The "first order" correction (in term of $d$, the distance between the 
two vertices) can be computed as in section \ref{sec:sim2V}, 
setting $\ft\sim \frac{dp}{2}$ in the above matrices. We focus 
on the case $\alpha_{2}=\eps\,\alpha_{1}$, and set 
$\beta=\alpha_{1}^2\,.$
% We get
% \begin{eqnarray}
% S(p) &\sim& \frac{-1/2}{
% \big({pd}+4i\beta\big)\,\big({pd}+i/\beta\big)} 
% \,\Big\{8\,S^{(0)}+i\frac{dp}{\beta}\,S^{(1)}
% -\left({dp}\right)^2\,S^{(2)}\Big\}\,,
% \quad\\[2.1ex]
% S^{(1)} &=& \left(\begin{array}{cccc} 
% -1 & 4\eps \beta^2 & 0 & 0 \\
% 4\eps \beta^2 & -1 & 0 & 0 \\ 
% 0 & 0 & -1 & 4\eps \beta^2 \\
% 0 & 0 & 4\eps \beta^2 & -1
% \end{array}\right)\ ;\ 
% S^{(2)} \ =\ \left(\begin{array}{cccc} 
% 1 & -\eps & 0 & 0 \\
% -\eps & 1 & 0 & 0 \\ 
% 0 & 0 & 1 & -\eps \\
% 0 & 0 & -\eps & 1
% \end{array}\right)\,.
% \end{eqnarray}
Multiplying by $U$, we get
 a first correction (in $d$) for the matrices given in 
 \cite{Cham2}:
\begin{eqnarray}
S^{(1)}(p) &\sim&  \frac{-1/2}{
\big({pd}+4i\beta\big)\,\big({pd}+i/\beta\big)}  
\,\Big\{8\,S^{(0)}_{ch}+i\frac{dp}{\beta}\,S^{(1)}_{ch}
-\left({dp}\right)^2\,S^{(2)}_{ch}\Big\}\,,
\label{eq:Schamtree}\\[2.1ex] 
S^{(1)}_{ch} &=& \left(\begin{array}{cccc} 
0 & 0 & 4\eps \beta^2 & -1 \\
0 & 0 & -1  & 4\eps \beta^2 \\ 
 4\eps \beta^2 & -1 & 0 & 0 \\
-1 & 4\eps \beta^2 & 0 & 0 
\end{array}\right)\ ;\
S^{(2)}_{ch} \ =\ \left(\begin{array}{cccc} 
0 & 0  & -\eps & 1\\
0 & 0 & 1 & -\eps \\ 
-\eps & 1 & 0 & 0\\
1 & -\eps &0 & 0 
\end{array}\right)\,.\qquad
\label{eq:Scham-cor}
\end{eqnarray}

\subsection{Example 2: the loop}
The loop graph has been treated in section \ref{sec:loop}. Using the form 
(\ref{eq:Sscal}), the total scattering matrix (\ref{eq:Sloop-gen}) rewrites 
\begin{eqnarray}
S_{tot}(p) &=& \frac{\exp(idp)}{\cN(p)}\,
\left(
\begin{array}{cc}
-2i\mu\,\sin(dp) & 1-\mu^{2} \\[1.2ex]
1-\mu^{2} & -2i\mu\,\sin(dp)
\end{array}
\right)
\mb{with} 
\mu=\frac{1-\alpha_{1}^2+\alpha_{2}^2}{1+\alpha_{1}^2+\alpha_{2}^2}  
\nonu
\cN(p) &=& 1-\big(\mu\,\exp(idp)\big)^2\,. 
\label{eq:Sloop-scale}
\end{eqnarray}
For scattering matrices obeying locally the Kirchhoff's rule, one has 
$\mu=\frac{1+\alpha_{1}}{1+\alpha_{1}+\alpha_{1}^2}$.

In the particular case $\mu=-\frac13$ (i.e. $\alpha_{1}=-2$ when 
Kirchhoff's rule is obeyed locally), one recovers the $S$ matrix 
found in example 3.2 of
\cite{Schra2}, with the identification $p\equiv \sqrt E$.

The inner modes take the form
\beq\begin{array}{l}
\displaystyle\fa_{2}(p) = \frac{\gamma\,\alpha_{2}}{\cN(p)}\,\Big(
e^{-ipd}\,\fa_{1}(-p)+\mu\,\fa_{3}(-p) 
\Big)\\[1.7ex]
\displaystyle\fa_{4}(p) = \frac{\gamma}{\cN(p)}\,\Big(
e^{-ipd}\,\fa_{1}(-p)+\mu\,\fa_{3}(-p)\Big) 
\end{array}\qquad
\gamma = \frac{2\alpha_{1}}{1+\alpha_{1}^2+\alpha_{2}^2}\,.
\eeq

\subsubsection{Expansion in term of the loop length}
We rewrite the scattering matrix in term of $\ft=\tan(dp/2)$:
\begin{eqnarray}
S_{tot}(p) &=& 
\left(\begin{array}{cc} \displaystyle
R(p)
&\displaystyle T(p)
\\[2.1ex]
\displaystyle T(p)
&\displaystyle R(p)
\end{array}\right)
\mb{with} \begin{cases} \displaystyle
R(p)= \frac{4i\mu\,\ft}
{(1-\mu^2)\ft^2+2i(1+\mu^2)\ft-1+\mu^2}\\[2.1ex]
\displaystyle T(p) = \frac{(\mu^2-1)(1+\ft^2)}
{(1-\mu^2)\ft^2+2i(1+\mu^2)\ft-1+\mu^2}
\end{cases}
\label{eq:Sloopshortexp}
\end{eqnarray}
When $d\to0$, we get an aproximation of the scattering matrix setting 
$\ft\sim dp/2$.
One recognizes in the approximation, 
the scattering matrix for a point-like impurity on the line. 
The reflection and transmission coefficients defining 
this impurity are given by the local parameters $\alpha_{1}$, 
$\alpha_{2}$, and by the distance $d$ (or equivalently by the 
surface $d^2$ of the loop). Correction to this approximation, induced 
by the surface of the loop, are given 
by the full expression (\ref{eq:Sloop-scale}).

\subsection{Example 3: the tadpole}
We apply the result for the tadpole graph (see 
section \ref{sec:tad}). 
Starting from the general form (\ref{eq:Sscal}), and using the 
expression (\ref{eq:Stot-tadpole}), this leads to the tadpole $S$ matrix
\begin{equation}
S_{tot}(p)=\left(
\begin{array}{cc} R(p) & 0 \\ 0 & 1 \end{array}
\right)
\mb{with}\displaystyle
R(p)=\frac{(1+\alpha_{1}^2+\alpha_{2}^2)\,e^{2idp} 
+4\alpha_2\,e^{idp}+1-\alpha_{1}^2+\alpha_{2}^2}
{(1-\alpha_{1}^2+\alpha_{2}^2)\,e^{2idp} 
+4\alpha_2\,e^{idp}+1+\alpha_{1}^2+\alpha_{2}^2}
\,.
 \label{eq:Rtad}
\end{equation}
We get a system with a half-line with reflection coefficient 1 decoupled 
from another half-line, with reflection coefficient $R(p)$.

In the particular case of $\alpha_{1}=\pm2$ and $\alpha_{2}=1$ 
 one recovers the $S$ 
matrix given in example 4.3 of \cite{Schra2}, again with identification 
$p\equiv \sqrt E$.

The modes inside the loop read 
\begin{eqnarray}
\fa_{2}(p) &=& 
\frac{-2\alpha_{1}(\alpha_{2}+e^{ipd})}{(1-\alpha_{1}^2+\alpha_{2}^2)\,e^{2idp} 
+4\alpha_2\,e^{idp}+1+\alpha_{1}^2+\alpha_{2}^2}\ \fa_{1}(-p)
\label{eq:a2-tadpol}\\
\fa_{4}(p) &=& 
\frac{-2\alpha_{1}(1+\alpha_{2}\,e^{ipd})}{(1-\alpha_{1}^2+\alpha_{2}^2)\,e^{2idp} 
+4\alpha_2\,e^{idp}+1+\alpha_{1}^2+\alpha_{2}^2}\ \fa_{1}(-p)
\label{eq:a4-tadpol}\,.
\end{eqnarray}

\subsubsection{Expansion in term of the loop length}
We rewrite the reflection coefficient (\ref{eq:Rtad}) as
\ben
R(p) &=& \frac{(1-\alpha_{2})^2\,\ft^2
-2i\alpha_{1}^2\,\ft-(1+\alpha_{2})^2}
{(1-\alpha_{2})^2\,\ft^2
+2i\alpha_{1}^2\,\ft-(1+\alpha_{2})^2} 
\mb{with} \ft=\tan(pd/2)
 \label{eq:Rtad-approx}
\een
and perform the short distance approximation 
setting $\ft\sim dp/2$. 

\subsection{Example 4: the triangle \label{sec:triang-scal}}
We present the calculation of the total scattering matrix for the 
triangle (as it has been explained in section \ref{sec:star-triang}), up 
to the end, 
for scale invariant local scattering matrices (\ref{eq:Sscal}).

To simplify the presentation, we consider three identical vertices 
with local scattering matrix (\ref{eq:Sscal}) with 
$\alpha_{1}=\alpha_{2}=1$:
$$
S_{0}=-\frac{1}{3}\,\left(\begin{array}{ccc}  
-1 & 2 & 2 \\ 2 & -1 & 2 \\ 2 & 2 & -1\end{array}\right)
$$
and 
take the same length $d$ for the three connecting lines.

At the first step of the gluing, we get a tree-graph matrix of the 
type (\ref{tree-scal}),
$$
S^{[12]}(p)=\frac{1}{9-e^{2idp}}
\left(\begin{array}{cccc} \displaystyle
e^{2idp}+3 & 2(e^{2idp}-3) & 4e^{idp} & 4e^{idp}\\
2(e^{2idp}-3) & e^{2idp}+3 & 4e^{idp} & 4e^{idp}\\
4e^{idp} & 4e^{idp} & e^{2idp}+3 & 2(e^{2idp}-3)\\ 
4e^{idp} & 4e^{idp} & 2(e^{2idp}-3) & e^{2idp}+3
\end{array}\right)
$$
that, after rotation by $P_{12}$, we glue to
$$
S^{[3]}(p) =-\frac{1}{3}\,\left(\begin{array}{ccc}  
-e^{2idp} & 2e^{2idp} & 2e^{idp} \\ 2e^{2idp} & -e^{2idp} & 2e^{idp} \\ 
2e^{idp} & 2e^{idp} & -1\end{array}\right)
$$
using the general formulas (\ref{eq:Stot-gen2V}).
We get
\begin{eqnarray}
S_{tot}(p) &=& \frac{3e^{ipd}+1}{e^{ipd}+3}\,\II_{3}
+\frac{4(e^{ipd}-1)e^{ipd}}{(e^{ipd}+3)(e^{2ipd}-2e^{ipd}+3)}\
\left(\begin{array}{ccc}
2 & -1 &-1 \\ -1 & 2 & -1 \\ -1 & -1 & 2
\end{array}\right)\,.\quad
\label{eq:Striangle-scal}
\end{eqnarray}
The inner modes $\fa_{4}(p)$ and $\fa_{6}(p)$ are obtained using 
formulas (\ref{eq:inner-gen}):
\begin{eqnarray*}
\fa_{4}(p) &=& \frac{2}{e^{2ipd}-2e^{ipd}+3}\,
\left\{
\frac{e^{ipd}-3}{e^{ipd}+3}\ \fa_{1}(-p)
+\frac{2e^{ipd}}{e^{ipd}+3}\ \fa_{3}(-p) 
-\frac{e^{ipd}\,(e^{ipd}+1)}{e^{ipd}+3}\ \fa_{5}(-p) 
\right\}
\\
\fa_{6}(p) &=& \frac{2}{e^{2ipd}-2e^{ipd}+3}\,
\left\{
\frac{2e^{ipd}}{e^{ipd}+3}\ \fa_{1}(-p)
+\frac{e^{ipd}-3}{e^{ipd}+3}\ \fa_{3}(-p) 
-\frac{e^{ipd}\,(e^{ipd}+1)}{e^{ipd}+3}\ \fa_{5}(-p) 
\right\}\,.
\end{eqnarray*}
 The mode $\fa_{2}(p)$ is obtained according to the calculation 
 explained in section \ref{sec:star-triang}:
\begin{eqnarray*}
\fa_{2}(p) &=& \frac{2}{e^{2idp}-9}\Big\{
3\, \fa_{1}(-p) 
+3\, \fa_{4}(-p) 
+e^{ipd}\,\fa_{6}(p) +e^{ipd}\,\fa_{3}(-p) \Big\}
\,.
\end{eqnarray*}

\subsubsection{Expansion in term of the distance}
An equivalent form of (\ref{eq:Striangle-scal}) is given by
\begin{eqnarray}
S_{tot}(p) &=& -\frac{\ft-2i}{\ft+2i}\,\II_{3}
+\frac{2\ft\,(\ft^2+1)}{(\ft+2i)(3\ft^2+2i\ft-1)}\
\left(\begin{array}{ccc}
2 & -1 &-1 \\ -1 & 2 & -1 \\ -1 & -1 & 2
\end{array}\right)\,.\quad
\label{eq:Striangle}
\end{eqnarray}
where $\ft=\tan(dp/2)$, the short distance expansion being given by 
$\ft\sim dp/2$.

One can compare this matrix with the symmetric scattering matrix 
introduced in \cite{Cham1}:
\begin{eqnarray}
\cR &=& \frac13
\left(\begin{array}{ccc}
-1 & 2 & 2 \\ 2 & -1 & 2 \\ 2 & 2 & -1
\end{array}\right) \,.
\end{eqnarray}
To get this matrix at $d=0$, we multiply $S_{tot}(p)$ by $\cR$. It is 
possible because we have $\cR^2=\II$ and $[S_{tot}(p)\,,\,\cR]=0$, so that 
$\cR\,S_{tot}(p)$ is still unitary. Then, the short distance 
approximation leads to:
\beq
\cR\,S_{tot}(p) \sim -\frac{\ft-2i}{\ft+2i}\,\cR
+\frac{2\ft\,(\ft^2+1)}{(\ft+2i)(3\ft^2+2i\ft-1)}\
\left(\begin{array}{ccc}
-2 & 1 & 1 \\ 1 & -2 & 1 \\ 1 & 1 & -2
\end{array}\right)\,,\quad\ft\sim\frac{dp}{2}\,,\quad
\label{eq:Schamstar}
\eeq
which can be viewed as a first correction to the scattering matrix 
$\cR$. 

\section{Conductance\label{sec:conduc}}
\subsection{General settings}
The conductance of a quantum wire is obtained through the linear 
response of the current $j_{\mu,a}(x,t)$ to a classical external potential 
$A_{\mu,a}(x,t)$ minimally coupled 
to a fermionic field $\psi_{b}(x,t)$ on the quantum wire. This has 
been treated in \cite{bosostar} for Tomonaga-Luttinger (TL) model in the case of star graphs, see also 
\cite{Schra-mag} for a general treatment of self-adjoint magnetic 
Laplacian on graphs. When considering a general 
quantum wire, we will restrict ourself to the problem of computing 
the conductance within the TL model and between external edges only.
Via bosonization, all the problem can be rewritten in term of the 
bosonic field $\phi_{a}(x,t)$ given in (\ref{eq:freefield}), 
see \cite{bosostar} for details.
 For instance, the local gauge transformations read
\ben
A_{\mu,a}(x,t) &\to& A_{\mu,a}(x,t)- \prt_{\mu}\,\Lambda_{a}(x,t) 
\mb{,} \mu=x\,,\,t\mb{;}a=1,\ldots,\fn\\
\phi_{a}(x,t) &\to& \phi_{a}(x,t)+ 
\frac{1}{\sigma\,\sqrt{\pi}}\,\Lambda_{a}(x,t) 
\een
and the corresponding invariant current reads
\beq
j_{\mu,a}(x,t) = \sqrt{\pi}\,\prt_{\mu}\,\phi_{a}(x,t)+ 
\frac{1}{\sigma}\,A_{\mu,a}(x,t)\,,
\eeq
 with possibly 
some additional terms corresponding to bound states \cite{boundstate}.
Let us stress that this current is just the bosonized version of a 
fermionic (relativistic) current $\psi(x,t)\gamma_{\mu}\psi(x,t)$.
Then, the linear response theory leads to
$$
<j_{x,a}(x,t)>_{A_{\mu}} = \frac{1}{\sigma}A_{x,a}(x,t)
+\frac{i}{\sigma}\sum_{b=1}^\fn \int_{-\infty}^t d\tau 
\int_{0}^\infty dy\ A_{x,b}(y,\tau)\,<[\prt_{x}\phi_{b}(y,\tau)\,,\, 
\prt_{x}\phi_{a}(x,t)]>
$$
Considering a uniform electric field switched on at $t=t_{0}$,
 in the Weyl gauge:
\ben
&& E_{a}(t) = \prt_{t}\,A_{x,a}(t) 
\mb{with} A_{x,a}(t)=0 \mb{if} t<t_{0} \\
&& A_{t,a} =0 \,,\ 
\forall t
\mb{,} a=1,\ldots,\fn
\een
and supposing that the scattering matrix is symmetric and admits  
simple non-real poles only, 
one can derive the conductance \cite{boundstate}
\ben
<j_{x,a}(x,t)>_{A_{\mu}} &=& \sum_{b=1}^\fn 
\int_{-\infty}^\infty \frac{d\omega}{2\pi}\,\wh 
A_{x,b}(\omega)\,G_{ab}(\omega,t-t_{0})\,e^{-i\omega t}\,,\\
G_{ab}(\omega,t) &=& G_{0}\,\Big\{
\delta_{ab}-S_{ab}(\omega)-\sum_{\eta\in\cP} 
\frac{e^{i(\omega-i\eta)t}}{\omega-i\eta}\,\cT_{ab}(\eta)\,
\Big\}\,.
\label{eq:cond}
\een
We have also introduced $\cP$, the set of poles of the scattering 
matrix and
\beq
\cT_{ab}(\eta)= \lim_{p\to\,i\eta} 
(p-i\eta)\,S_{ab}(p)\,.
\eeq
$G_{0}$ is the conductance for the infinite line.
The conductance depends on the time $t_{0}$ of the switch-on of the electric 
field, but also (due to the presence of poles in the scattering 
matrix) on its frequency $\omega$. 
\paragraph{Short distance approximation for the conductance:}
The formula (\ref{eq:cond}) can be 
applied to any of the scattering matrices computed in the previous 
sections, in particular to the ones deduced in the short distance 
approximation, where the number of poles is finite. Moreover, when 
performing this short distance approximation, and since the pole 
content has been already (partially) kept in the $\cT_{ab}$ matrix, one can take 
for the total scattering matrix in (\ref{eq:cond}) its value for $d=0$. 
Since in our examples the local scattering matrices are constant, this is equivalent to 
take $p=0$ in $S$, so that we get for the conductance the approximated 
form
\beq
G_{ab}^{approx}(\omega,t) = G_{0}\,\Big\{
\delta_{ab}-S_{ab}(0)-\sum_{\eta\in\cP_{0}} 
\frac{e^{i(\omega-i\eta)t}}{\omega-i\eta}\,\cT_{ab}(\eta)\,
\Big\}\,,
\label{eq:cond-approx}
\eeq
where $\cP_{0}$ is the set of poles appearing in the approximated scattering 
matrix of the graph under consideration. An obvious refinement of 
this approximation is to take 
\beq
G_{ab}^{refin}(\omega,t) = G_{0}\,\Big\{
\delta_{ab}-S_{ab}^{approx}(\omega)-\sum_{\eta\in\cP_{0}} 
\frac{e^{i(\omega-i\eta)t}}{\omega-i\eta}\,\cT_{ab}(\eta)\,
\Big\}\,,
\label{eq:cond-refin}
\eeq
 where $S_{ab}^{approx}(\omega)$ is the short distance 
approximation of the scattering matrix.

Some examples of such calculations are done in the next section.

\subsection{Examples}
We apply the above formalism to the examples dealt with in 
section \ref{sec:scalinv}. Except for one particular case, we will consider the scattering matrix 
within the short distance approximation, as it has been presented in 
section \ref{sec:shortexp}. In the case of the tadpole, we perform the 
exact calculation and show that the short distance approximation is in 
accordance with the exact result, justifying in this way the 
approximation. 

\subsubsection{Tree graph}
We start with the matrix (\ref{eq:Schamtree}), which possesses two 
simple poles
\beq
i\eta_{1} = \frac{-i}{d}\,4\beta
\mb{and} i\eta_{2} = \frac{i}{d\,\beta}\,.
\eeq
We recall that $\beta=\alpha_{1}^2$. One gets, using notations 
(\ref{eq:Scham-cor}):
\ben
\cT_{1} &=& \cT(\eta_{1})\ =\ \frac{-2i\beta}{4\beta^2-1}\Big\{
2\,S^{(0)}_{ch}+S^{(1)}_{ch}+4\beta^2\,S^{(2)}_{ch}
\Big\}\\[1.2ex]
\cT_{2} &=& \cT(\eta_{2})\ =\ \frac{i\beta}{4\beta^2-1}\Big\{
4\,S^{(0)}_{ch}+\half\,S^{(1)}_{ch}+\frac{1}{2\beta^2}\,S^{(2)}_{ch}
\Big\}
\quad
\een
Starting from the formula (\ref{eq:cond-approx}), it leads to a conductance
\ben
G(\omega,t)= G_{0}\,\Big\{ \half\left(\begin{array}{cc}
A & -A \\  -A & A
\end{array}\right)
-\frac{e^{i(\omega-i\eta_{1})t}}{\omega-i\eta_{1}}\,
\,\cT_{1}
-\frac{e^{i(\omega-i\eta_{2})t}}{\omega-i\eta_{2}}\,
\,\cT_{2}\Big\}\,,
\een
where we have introduced the $2\times2$ matrix
\beq
A=\left(\begin{array}{cc} 1 & \eps \\ \eps & 1 \end{array}\right)
\,.
\eeq

\subsubsection{Loop}
We consider the loop scattering matrix (\ref{eq:Sloopshortexp}). It 
possesses two simple poles:
\beq
i\eta_{1}=\frac{2i(\mu-1)}{d(\mu+1)}
=\frac{-2i\alpha_{1}^2}{(1+\alpha_{2}^2)\,d}
\mb{and}
i\eta_{2}=\frac{2i(\mu+1)}{d(\mu-1)}
=\frac{-2i(1+\alpha_{2}^2)}{\alpha_{1}^2\,d}\,.
\eeq
They lead to the two matrices
\beq
\cT_{1} = \cT(\eta_{1}) = i\frac{1+\mu}{1-\mu}\,
\left(\begin{array}{cc}
1 & -1 \\ -1 & 1
\end{array}\right) \mb{and}
\cT_{2} = \cT(\eta_{2}) = -i\frac{1-\mu}{1+\mu}\,
\left(\begin{array}{cc}
1 &1 \\ 1 & 1
\end{array}\right)\,.
\eeq
Then, the conductance (\ref{eq:cond-approx}) rewrites
\ben
G(\omega,t)= G_{0}\,\Big\{ \left(\begin{array}{cc}
1 & -1 \\  -1 & 1
\end{array}\right)
-\frac{e^{i(\omega-i\eta_{1})t}}{\omega-i\eta_{1}}\,
\,\cT_{1}
-\frac{e^{i(\omega-i\eta_{2})t}}{\omega-i\eta_{2}}\,
\,\cT_{2}\Big\}\,.
\een

\subsubsection{Tadpole}
To simplify the presentation, we consider the case $\alpha_{2}=1$, 
but the same sort of calculation 
can be done for the general case. 
\paragraph{Exact calculation:} 
The expression (\ref{eq:Rtad}) for $\alpha_{2}=1$ simplifies to
\beq
R(p) = \frac{2+i\alpha_{1}^2\,\ft}{2-i\alpha_{1}^2\,\ft}
\eeq
 The poles of $R(p)$ are given by:
$$
\tan(\frac{dp}{2})=\frac{-2i}{\alpha_{1}^2}
\quad\Leftrightarrow \quad p=i\eta_{k}=i\eta_{0}+\frac{2k\pi}{d}\,,\ k\in\ZZ
\mb{with} \eta_{0}=-\frac{2}{d}\,\arctanh(\frac{2}{\alpha_{1}^2})\,,
$$
that leads to
\beq
\cT_{k}=\cT(\eta_{k})=\frac{8i}{d}\,
\frac{\alpha_{1}^2}{\alpha_{1}^4-4}\,,\quad \forall 
k\in\ZZ\,.
\eeq
Hence, we get
\beq
G(\omega,t)=G_{0}\,\Big\{ 1-R(\omega)-\frac{8i}{d}\,
\frac{\alpha_{1}^2}{\alpha_{1}^4-4}\,\sum_{k\in\ZZ}
\frac{e^{i(\omega-i\eta_{k})t}}{\omega-i\eta_{k}}\Big\}\,.
\eeq
The sum can be computed for real parameter $\alpha_{1}$, and one obtains
\beq
\sum_{k\in\ZZ}
\frac{1}{\omega-i\eta_{k}}\,e^{i(\omega-i\eta_{k})t}=
\frac{-i\,d\,e^{i\,n_{0}(t)\,d\,(\omega-i\eta_{0})}}{1-e^{i\,d\,(\omega-i\eta_{0})}}
\mb{with} n_{0}(t)=\left[\frac{t}{d}\right]
\eeq
where $[.]$ denotes the integer part. This leads to
\beq
G(\omega,t)=G_{0}\,\Big\{ 1-\frac{2 +i\,\alpha_{1}^2\,\tan(\frac{\omega d}{2})}
{2 -i\,\alpha_{1}^2\,\tan(\frac{\omega d}{2})}-
\frac{8\alpha_{1}^2}{\alpha_{1}^4-4}\,
\frac{e^{i\,n_{0}(t)\,d\,(\omega-i\eta_{0})}}
{1-e^{id(\omega-i\eta_{0})}}\Big\}\,.
\label{eq:tadpol-exact}
\eeq
\paragraph{Short distance approximation:} If one performs the same 
calculation with the  
approximation $\ft\sim pd/2$, we get a single simple pole
$i\eta'={-4i}/{d\,\alpha_{1}^2}$ that leads to
\beq
G_{approx}(\omega,t)=G_{0}\,\Big\{ 1-\frac{2 +i\,\alpha_{1}^2\,\frac{\omega d}{2}}
{2 -i\,\alpha_{1}^2\,\frac{\omega d}{2}}+
\frac{8i}{d\,\alpha_{1}^2}\,
\frac{e^{i(\omega-i\eta')t}}{\omega-i\eta'}\Big\}\,.
\eeq
To compare this latter expression with the exact result, we first 
note that 
\beq
\eta_{0} = -\frac{2}{d}\,\arctanh(\frac{2}{\alpha_{1}^2})
\sim -\frac{4}{d\alpha_{1}^2}=\eta' \mb{for} \alpha_{1}^2>>2
\eeq
Taking this regime for the parameter $\alpha_{1}$,
we perform an 
expansion in $d$ of (\ref{eq:tadpol-exact}), remarking that 
$n_{0}(t)\,d\sim t$ when $d\to0$:
\beq
G(\omega,t)\sim G_{0}\,\Big\{ 1-\frac{2 +i\,\alpha_{1}^2\,\frac{\omega d}{2}}
{2 -i\,\alpha_{1}^2\,\frac{\omega d}{2}}+
\frac{8}{\alpha_{1}^2}\ 
\frac{e^{i(\omega-i\eta')t}}{(-id)(\omega-i\eta')}\Big\}\,,
% \mb{for} d\to0
\eeq
which is exactly the expression of $G_{approx}(\omega,t)$. Thus, the 
short distance approximation gives a correct estimate of the 
conductance for this parameter range. 

\subsubsection{Triangle}
The matrix (\ref{eq:Schamstar}) possesses three 
simple poles 
\beq
i\eta_{0}=\frac{-4i}{d}\mb{;} 
i\eta_{\pm}=\frac{-2i\pm2\sqrt{2}}{3\,d}\,,
\eeq
leading to
\ben
\cT(\eta_{0})=\frac{8i}{3d}\,\left(\begin{array}{ccc}
1 & 1 & 1 \\ 1 & 1 & 1 \\ 1 & 1 & 1
\end{array}\right)
\mb{and}
\cT(\eta_{\pm})=-\frac{2}{9d}(2i\pm\sqrt{2})\,\left(\begin{array}{ccc}
2 & -1 & -1 \\ -1 & 2 & -1 \\ -1& -1& 2
\end{array}\right)\,.
\een
The conductance takes the form
\ben
G(\omega,t)= G_{0}\,\Big\{\II_{3}-\cR
-\frac{e^{i(\omega-i\eta_{0})t}}{\omega-i\eta_{0}}\,\cT(\eta_{0})
-\frac{e^{i(\omega-i\eta_{+})t}}{\omega-i\eta_{+}}\,\cT(\eta_{+})
-\frac{e^{i(\omega-i\eta_{-})t}}{\omega-i\eta_{-}}\,\cT(\eta_{-})
\Big\}\,.
\een
When the distance $d\to0$, the $\omega$ dependent part of the 
conductance goes to zero, and one recover the conductance computed 
in \cite{Cham2}. For non-vanishing values of $d$, we get $\omega$ dependent 
corrections to this conductance. 
 
\section*{Acknowledgements}
It is a pleasure to warmly thank M. Mintchev for many fruitful 
discussions and comments, and for providing me the reference \cite{Schra-unp}.
I am grateful to  V. Caudrelier for his useful remarks on the manuscript.
I wish to thank the referee for his useful coments and suggestions.

\appendix
\section{Proofs\label{app:proof}}
\subsection{Compatibility relation\label{app:cons}}
We start with the relations at each vertex
\ben
A_{1}(p) &=& S_{11}(p)\,A_{1}(-p) +S_{12}(p)\,A_{2}(-p) 
\label{eq:a1}\\
A_{2}(p) &=& S_{21}(p)\,A_{1}(-p) +S_{22}(p)\,A_{2}(-p) 
\label{eq:a2}\\
A_{2}(-p) &=& \Sigma_{11}(p)\,A_{2}(p) +\Sigma_{12}(p)\,A_{3}(-p) 
\label{eq:a2bis}\\
A_{3}(p) &=& \Sigma_{21}(p)\,A_{2}(p) +\Sigma_{22}(p)\,A_{3}(-p) 
\label{eq:a3}
\een
where we used the notations of section \ref{sec:glcal}. Equations (\ref{eq:a2}) 
and (\ref{eq:a2bis}) allow to express $A_{2}(p)$ in term of $A_{1}(p)$ 
and $A_{3}(p)$ in two different ways:
\ben
A_{2}(p) &=& \wt D(p)^{-1}\,\Big(S_{21}(p)\,A_{1}(-p) 
+S_{22}(p)\,\Sigma_{12}(p)\,A_{3}(-p) \Big)
\label{eq:sola2}\\
A_{2}(-p) &=& D(p)^{-1}\,\Big(\Sigma_{12}(p)\,A_{3}(-p) 
+\Sigma_{11}(p)\,S_{21}(p)\,A_{1}(-p) \Big)
\label{eq:sola2bis}\\
\mb{with}&&  \wt 
D(p)=\II-S_{22}(p)\,\Sigma_{11}(p) \mb{and}
D(p)=\II-\Sigma_{11}(p)\,S_{22}(p)
\een
Plugging (\ref{eq:sola2}) into (\ref{eq:a3}) and (\ref{eq:sola2bis}) into (\ref{eq:a1})
leads to the relations (\ref{eq:boundStot})-(\ref{eq:Stot-gen2V}). 
(\ref{eq:sola2}) can be viewed as 
the determination of $A_{2}(p)$ in terms of $A_{1}(p)$ 
and $A_{3}(p)$. It remains a compatibility relation:
\ben
&&\wt D(p)^{-1}\,\Big(S_{21}(p)\,A_{1}(-p) 
+S_{22}(p)\,\Sigma_{12}(p)\,A_{3}(-p)\Big)
\ =\ \nonu
&&\ =\  D(-p)^{-1}\,\Big(\Sigma_{12}(-p)\,A_{3}(p) 
+\Sigma_{11}(-p)\,S_{21}(-p)\,A_{1}(p)\Big)
\een
which rewrites, using again relations (\ref{eq:boundStot})-(\ref{eq:Stot-gen2V}),
\ben
D(-p)\,\wt D(p)^{-1}\,\Big(S_{21}(p)\,A_{1}(-p) 
+S_{22}(p)\,\Sigma_{12}(p)\,A_{3}(-p)\Big)\ =\ 
\Big(\Sigma_{12}(-p)\,\Sigma_{21}(p)\,\wt D(p)^{-1}
\qquad\nonu
-\Sigma_{11}(-p)\,S_{22}(-p) +\Sigma_{11}(-p)\,S_{21}(-p)\,S_{12}(p)
\,D(p)^{-1}\,\Sigma_{11}(-p)\Big)S_{21}(p)\,A_{1}(-p)
\qquad\nonu
+\Big(\Sigma_{11}(-p)\,S_{21}(-p)\,S_{12}(p)\,D(p)^{-1}
-\Sigma_{11}(-p) +\Sigma_{12}(-p)\,\Sigma_{21}(p)\,\wt D(p)^{-1}\,S_{22}(p)
\Big)\Sigma_{12}(p)\,A_{3}(-p) 
\nonumber
\een
Instead of proving this relation, we prove the two following relations, that 
obviously imply 
the compatibility relation,
\ben
D(-p)\,\wt D(p)^{-1}&=& 
\Sigma_{12}(-p)\,\Sigma_{21}(p)\,\wt D(p)^{-1}
 +\Sigma_{11}(-p)\,S_{21}(-p)\,S_{12}(p)
\,D(p)^{-1}\,\Sigma_{11}(p)
\nonu
&&-\Sigma_{11}(-p)\,S_{22}(-p)
\label{eq:a}\\
D(-p)\,\wt D(p)^{-1}\,S_{22}(p)&=&
\Sigma_{11}(-p)\,S_{21}(-p)\,S_{12}(p)\,D(p)^{-1}
 +\Sigma_{12}(-p)\,\Sigma_{21}(p)\,\wt D(p)^{-1}\,S_{22}(p)
\nonu
&&-\Sigma_{11}(-p)\label{eq:b}
\een
We start by proving relation (\ref{eq:a}). Multiplying on the right 
by $\wt D(p)$ and using the consistency relation (\ref{cons}) for 
$S(p)$ and $\Sigma(p)$, it can be rewritten as
\ben
D(-p)&=& \II-\Sigma_{11}(-p)\Sigma_{11}(p)
 +\Sigma_{11}(-p)\big(\II-S_{22}(-p)S_{22}(p)\big)
D(p)^{-1}\Sigma_{11}(p)\,\wt D(p)
\nonu
&&-\Sigma_{11}(-p)\,S_{22}(-p)\big(\II-S_{22}(p)\Sigma_{11}(p)\big)
\een
that is indeed an equality.
Relation (\ref{eq:b}) is equivalent to relation (\ref{eq:a}) 
multiplied from the right by $S_{22}(p)$.

\subsection{Consistency and Hermitian analycity relations}
We prove that the scattering matrix (\ref{eq:Stot-gen2V}) obeys the consistency and 
Hermitian analycity relations (\ref{cons}) and (\ref{herm}) as soon as 
the local scattering matrices do. 
The proof relies on the relations (proven by direct calculation):
\ben 
&& \Sigma_{11}(p)\,\wt D(p)^{-1} = D(p)^{-1}\,\Sigma_{11}(p)
\mb{and} \wt D(p)^{-1}\,S_{22}(p) = S_{22}(p)\,D(p)^{-1}
\label{eq:DDt}\\
&& \Big(\wt D(p)\Big)^\dag = D(-p)
\mb{and} \Big(D(p)\Big)^\dag = \wt D(-p)
\een
Thanks to these relations, it is easy to show that $S_{tot}(p)$ is 
Hermitian analytical. For instance one has
\ben
\Big(S_{tot}(p)\Big)^\dag_{11} &=& S_{11}(p)^\dag + S_{21}(p)^\dag
\Sigma_{11}(p)^\dag\Big(D(p)^{-1}\Big)^\dag S_{12}(p)^\dag
\nonu
&=& S_{11}(-p) + S_{21}(-p)
\Sigma_{11}(-p)\wt D(-p)^{-1} S_{12}(-p)
=\Big(S_{tot}(-p)\Big)_{11}
\nonumber
\een
The proof of the consistency relation requires more calculation. 
Considering the 11 component, and using consistency relations for 
$S(p)$ and $\Sigma(p)$, one can rewrites it as:
\ben
\Big(S_{tot}(p)\,S_{tot}(-p)\Big)_{11} &=& 
S_{11}(p)\,S_{11}(-p)+S_{12}(p)\,D(p)^{-1}\,\Big\{\ldots\Big\}\,S_{21}(-p)
\nonu
\Big\{\ldots\Big\} &=& D(p) +\Big[\ldots\Big]\,\wt D(-p)^{-1}
\nonu
\Big[\ldots\Big] &=&
\Big(\Sigma_{11}(p)-S_{22}(-p)\Big)\,D(-p)^{-1}\,\Big(\ldots\Big) \nonu
\Big(\ldots\Big) &=& -D(-p)\,\Sigma_{11}(-p)
+\Sigma_{11}(-p)\,\wt D(-p)
\ =\ 0
\een
where in the last step we used (\ref{eq:DDt}). 

The other relations are 
proven along the same lines.
%\section{biblio}

\end{document}